\documentclass[11pt]{article}
\usepackage{graphicx}
\usepackage{amsfonts}
\usepackage{amssymb}
\usepackage{url}

\newcommand{\hide}[1]{}
\newcommand\R{\mathbb{R}}
\newcommand{\squeezelist}{\setlength{\itemsep}{0pt}}
\newtheorem{theorem}{{\bf Theorem}}

\newtheorem{lemma}[theorem]{Lemma}
\newcommand{\ABox}{
\raisebox{3pt}{\framebox[6pt]{\rule{6pt}{0pt}}}
}
\newenvironment{proof}{{\bf Proof:}}{\hfill\ABox}

\def\C{{\cal C}}

{\makeatletter
 \gdef\xxxmark{%
   \expandafter\ifx\csname @mpargs\endcsname\relax 
     \expandafter\ifx\csname @captype\endcsname\relax 
       \marginpar{xxx}
     \else
       xxx 
     \fi
   \else
     xxx 
   \fi}
 \gdef\xxx{\@ifnextchar[\xxx@lab\xxx@nolab}
 \long\gdef\xxx@lab[#1]#2{{\bf [\xxxmark #2 ---{\sc #1}]}}
 \long\gdef\xxx@nolab#1{{\bf [\xxxmark #1]}}
 \gdef\turnoffxxx{\long\gdef\xxx@lab[##1]##2{}\long\gdef\xxx@nolab##1{}}%
}

\begin{document}

\title{Grid Vertex-Unfolding Orthogonal Polyhedra
\thanks{This is a significant revision of the preliminary version that
appeared in~\protect\cite{dfo-gvuop-06}.}
}

%
%
%
%
%

\author{
Mirela Damian%
   \thanks{Dept. Comput. Sci., Villanova Univ., Villanova,
    PA 19085, USA.
   \protect\url{mirela.damian@villanova.edu}.}
\and
Robin Flatland%
   \thanks{Dept. Comput. Sci., Siena College, Loudonville, NY 12211, USA.
    \protect\url{flatland@siena.edu}.}
\and
Joseph O'Rourke%
    \thanks{Dept. Comput. Sci., Smith College, Northampton, MA
      01063, USA.
      \protect\url{orourke@cs.smith.edu}.
       Supported by NSF
       award
       DUE-0123154.}
}

\maketitle              

\begin{abstract}
An {\em edge-unfolding} of a polyhedron is produced by cutting along
edges and flattening the faces to a {\em net}, a connected planar
piece with no overlaps. A {\em grid unfolding} allows additional
cuts along grid edges induced by coordinate planes passing through
every vertex. A {\em vertex-unfolding} permits faces in the net to
be connected at single vertices, not necessarily along edges.  We
show that any orthogonal polyhedra of genus zero has a grid
vertex-unfolding. (There are orthogonal polyhedra that cannot be
vertex-unfolded, so some type of ``gridding'' of the faces is
necessary.) For any orthogonal polyhedron $P$ with $n$ vertices, we
describe an algorithm that vertex-unfolds $P$ in $O(n^2)$ time.
Enroute to explaining this algorithm, we present a simpler
vertex-unfolding algorithm that requires a $3 \times 1$ refinement
of the vertex grid.
\end{abstract}


\section{Introduction}
Two unfolding problems have remained unsolved for many
years~\cite{do-sfucg-05}: (1)~Can every convex polyhedron be
edge-unfolded? (2)~Can every polyhedron be unfolded? An
\emph{unfolding} of a 3D object is an isometric mapping of its
surface to a single, connected planar piece, the ``net'' for the
object, that avoids overlap. An \emph{edge-unfolding} achieves the
unfolding by cutting edges of a polyhedron, whereas a \emph{general
unfolding} places no restriction on the cuts.
A net representation of a polyhedron finds use in a variety of
applications~\cite{o-fucg-00} --- from flattening monkey
brains~\cite{ssw-nsgmm-89} to manufacturing from sheet
metal~\cite{w-mddsmp-97} to low-distortion texture
mapping~\cite{thcm-pm-04}.

It is known that some nonconvex polyhedra cannot be unfolded without
overlap with cuts along edges. However, no example is known of a
nonconvex polyhedron that cannot be unfolded with unrestricted cuts.
Advances on these difficult problems have been made by specializing
the class of polyhedra, or easing the stringency of the unfolding
criteria. On one hand, it was established
in~\cite{bddloorw-uscop-98} that certain subclasses of
\emph{orthogonal polyhedra} --- those whose faces meet at angles
that are multiples of $90^\circ$ --- have an unfolding. In
particular, the class of \emph{orthostacks}, stacks of extruded
orthogonal polygons, was proven to have an unfolding (but not an
edge-unfolding). On the other hand, loosening the criteria of what
constitutes a net to permit connection through points/vertices, the
so-called \emph{vertex-unfoldings}, led to an algorithm to
vertex-unfold any triangulated manifold~\cite{deeho-vusm-03} (and
indeed, any simplicial manifold in higher dimensions).
A vertex unfolding maps the surface to a single, connected piece $P$
in the plane, but $P$ may have ``cut vertices'' whose removal
disconnect $P$.

A second loosening of the criteria is the notion of grid unfoldings,
which are especially natural for orthogonal polyhedra. A \emph{grid
unfolding} adds edges to the surface by intersecting the polyhedron
with planes parallel to Cartesian coordinate planes through every
vertex. The two approaches were recently married
in~\cite{dil-gvuo-04}, which established that any orthostack may be
grid vertex-unfolded. For orthogonal polyhedra, a grid unfolding is
a natural median between edge-unfoldings and unrestricted
unfoldings.

Our main result is that any orthogonal polyhedron, without shape
restriction except that its surface be homeomorphic to a sphere, has
a grid vertex-unfolding. We present an algorithm that grid
vertex-unfolds any orthogonal polyhedron with $n$ vertices in
$O(n^2)$ time. We also present, along the way, a simpler algorithm
for $3 \times 1$ \emph{refinement} unfolding, a weakening of grid
unfolding that we define below. We believe that the techniques in
our algorithms may help show that all orthogonal polyhedra can be
grid edge-unfolded.

\section{Definitions}
\label{sec:defs}
A $k_1 \times k_2$ \emph{refinement} of a surface~\cite{do-op04-05}
partitions each face into a $k_1 \times k_2$ grid of faces. We will
consider refinements of grid unfoldings, with the convention that a
$1 \times 1$ refinement is an unrefined grid unfolding.

We distinguish between a \emph{strict net}, in which the net
boundary does not self-touch, and a \emph{net} for which the
boundary may touch, but no interior points overlap. The latter
corresponds to the physical model of cutting out the net from a
sheet of paper, with perhaps some cuts representing \emph{edge
overlap}, and this is the model we use in this paper.
We also insist as part of the definition of a vertex-unfolding,
again keeping in spirit with the physical model, that the unfolding
``path'' never self-crosses on the surface in the following sense.
If $(A,B,C,D)$ are four faces incident in that cyclic order to a
common vertex $v$, then the net does not include both the
connections $AvC$ and $BvD$.\footnote{
   This was not part of the original definition in~\cite{deeho-vusm-03} but was
   achieved by those unfoldings.
}

We use the following notation to describe the six type of faces of
an orthogonal polyhedron, depending on the direction in which the
outward normal points: {\em front}: $-y$; {\em back}: $+y$; {\em
left}: $-x$; {\em right}: $+x$; {\em bottom}: $-z$; {\em top}: $+z$.
We take the $z$-axis to define the vertical direction;
\emph{vertical} faces are parallel to the $xz$-plane. Directions
clockwise (cw), and counterclockwise (ccw) are defined from the
perspective of a viewer positioned at $y = -\infty$. We distinguish
between an original vertex of the polyhedron, which we call a
\emph{corner vertex} or just a \emph{vertex}, and a
\emph{gridpoint}, a vertex of the grid (which might be an original
vertex). A {\em gridedge} is an edge segment with both endpoint
gridpoints, and a {\em gridface} is a face of the grid bounded by
gridedges.

\begin{figure}[htbp]
\centering
\begin{tabular}{c@{\hspace{0.2\linewidth}}c}
\includegraphics[width=0.36\linewidth]{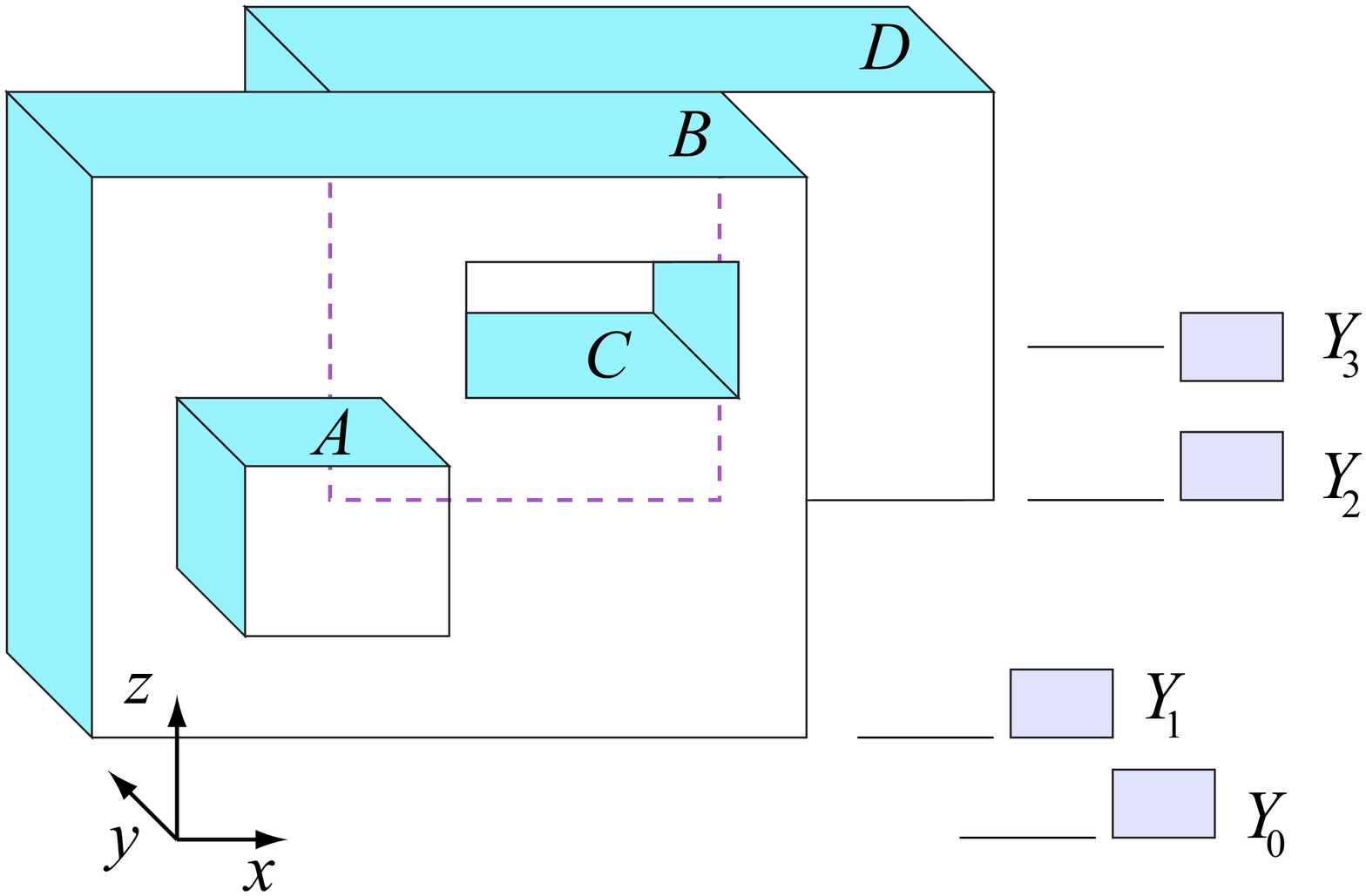} &
\raisebox{2em}{\includegraphics[width=0.08\linewidth]{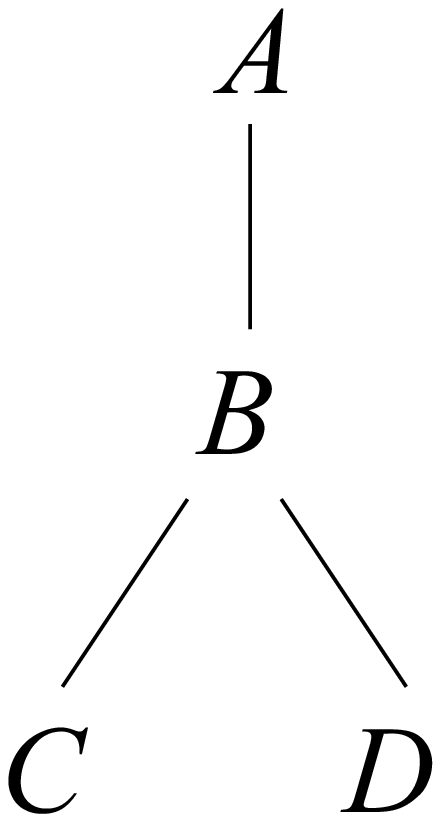}}
\end{tabular}
\caption{Definitions. (a) Shaded connected pieces are bands; $A$,
$B$ and $D$ are protrusions; $C$ is a dent; $r_2(CB)$ coincides with
the back face of $C$; $R_2(DB)$ is marked in dashed lines. (b) The
adjacency structure of bands is a tree. \label{fig:defs} }
\end{figure}

Let $O$ be a solid orthogonal polyhedron with the surface
homeomorphic to a sphere (i.e, genus zero). Let $Y_i$ be the plane
$y=y_i$ orthogonal to the $y$-axis. Let $Y_0, Y_1, \ldots, Y_i,
\ldots$ be a finite sequence of parallel planes passing through
every vertex of $O$, with $y_0 < y_1 < \cdots < y_i < \cdots$. We
define \emph{layer $i$} to be the portion of $O$ between planes
$Y_i$ and $Y_{i+1}$. Observe that a layer may include a collection
of disjoint connected components of $O$; we call each such component
a {\em slab}. A surface piece that surrounds a slab is called a {\em
band}. Referring to Fig.~\ref{fig:defs}a, layer $0$, $1$ and $2$
each contain one slab (with outer bands $A$, $B$ and $D$,
respectively). Note that each slab is bounded by an outer (surface)
band, but it may also contain inner bands, bounding holes. Outer
bands are called \emph{protrusions} and inner bands are called
\emph{dents} ($C$ in Fig.~\ref{fig:defs}a). In other words, band $A$
is a \emph{protrusion} if a traversal of the rim of $A$ in $Y_i$,
ccw from the viewpoint of $y=-\infty$, has the interior of $O$ to
the left of $A$, and a \emph{dent} if this traversal has the
interior of $O$ to the right.

For fixed $i$, define $P = \partial O \cap Y_i$ as the portion of
the surface of $O$ lying in plane $Y_i$. $P^+$ is the portion of $P$
with normal in the direction $+y$ (composed of back faces), and
$P^-$ the portion with normal in the direction $-y$ (composed of
front faces). By convention, band points in $P$ that are not
incident to either front or back faces (e.g., when one band aligns
with another), belong to both $P^+$ and $P^-$. Thus $P = P^+ \cup
P^-$.

For a band $A$, Let $R_i(A) = A \cap Y_i$ be the polygon in $Y_i$
determined by the rim of band $A$, and $r_i(A)$ the closed region of
$Y_i$ whose boundary is $R_i(A)$. For any two bands $A$ and $B$, let
$r_i(AB) = r_i(A) \cap r_i (B)$ and let $R_i(AB)$ be the boundary of
$r_i(AB)$.

\section{Dents vs. Protrusions}
We observe that dents may be treated exactly the same as protrusions
with respect to unfolding, because an unfolding of a $2$-manifold to
another surface (in our case, a plane) depends only on the intrinsic
geometry of the surface, and not on how it is embedded in $\R^3$.
Note that we are only concerned with the final unfolded ``flat
state''~\cite{do-sfucg-05}, and not with possible intersections
during a continuous sequence of partially unfolded intermediate
states. Our unfolding algorithm relies solely on the amount of
surface material surrounding each point: the cyclic ordering of the
faces incident to a vertex, and the pair of faces sharing an edge.
All these local relationships remain unchanged if we conceptually
``pop-out'' dents to become protrusions, i.e., a ``Flatland''
creature living in the surface could not tell the difference; nor
can our algorithm. We note that the popping-out is conceptual only,
for it could produce self-intersecting objects. Also dents are
gridded independently of the rest of the object, so that it would
not matter whether they are popped out or not.
%

Although the dent/protrusion distinction is irrelevant to the
unfolding, the interrelationships between dents and protrusions
touching a particular $Y_i$ do depend on this distinction. To cite
just the simplest example, there cannot be two nested protrusions to
the same side of $Y_i$, but a protrusion could have a dent in it to
the same side of $Y_i$ (e.g., protrusion $B$ encloses dent $C$ to
the same side of $Y_1$ in Fig.~\ref{fig:defs}). These relationships
are crucial to the connectivity of the band graph $G_b$, discussed
in Sec.~\ref{sec:Gb.connected}.

\section{Overview}
\label{sec:overview} The two algorithms we present share a common
central structure, with the second achieving a stronger result; both
are vertex-unfoldings that use orthogonal cuts only. We note that it
is the restriction to orthogonal cuts that makes the
vertex-unfolding problem difficult: if arbitrary cuts are allowed,
then a general vertex-unfolding can be obtained by simply
triangulating each face and applying the algorithm
from~\cite{deeho-vusm-03}.

The ($3 \times 1$)-algorithm unfolds any genus-0 orthogonal
polyhedron that has been refined in one direction 3-fold. The bands
themselves are never split (unlike in~\cite{bddloorw-uscop-98}). The
algorithm is simple. The ($1 \times 1$)-algorithm also unfolds any
genus-0 orthogonal polyhedron, but this time achieving a grid
vertex-unfolding, i.e., without refinement. This algorithm is more
delicate, with several cases not present in the ($3 \times
1$)-algorithm that need careful detailing. Clearly this latter
algorithm is stronger, and we vary the detail of presentation to
favor it. The overall structure of the two algorithms is the same:
\begin{enumerate}
\squeezelist
\item A band ``unfolding tree'' $T_U$ is constructed by shooting rays vertically
from the top of bands. The root of $T_U$ is a {\em frontmost} band
(of smallest $y$-coordinate), with ties broken arbitrarily.
\item A forward and return \emph{connecting} path of vertical faces
is identified, each of which connects a parent band to a child band
in $T_U$.
\item Each band is unfolded horizontally as a unit, but interrupted
when a connecting path to a
child is encountered.  The parent band unfolding is suspended at that
point, and the child band is unfolded recursively.
\item The vertical front and back faces of each slab are partitioned
according to an illumination model, with variations for the more
complex ($1 \times 1$)-algorithm. These vertical faces are attached
below and above appropriate horizontal sections of the band
unfolding.
\end{enumerate}
The final unfolding lays out all bands horizontally, with the
vertical faces hanging below and above the bands.  Non-overlap is
guaranteed by this strict two-direction structure.

Although our result is a broadening of that in~\cite{dil-gvuo-04}
from orthostacks to all orthogonal polyhedra, we found it necessary
to employ techniques different from those used in that work. The
main reason is that, in an orthostack, the adjacency structure of
bands yields a path, which allows the unfolding to proceed from one
band to the next along this path, never needing to return. In an
orthogonal polyhedron, the adjacency structure of bands yields a
tree (cf. Fig.~\ref{fig:defs}b). Thus unfolding band-by-band leads
to a tree traversal, which requires traversing each arc of the tree
in both directions. It is this aspect which we consider our main
novelty, and which leads us to hope for an extension to
edge-unfoldings as well.

\section{$(3 \times 1)$-Algorithm}


\subsection{Computing the Unfolding Tree $T_U$}
\label{sec:3x1tree}


Define a {\em z-beam} to be a vertical rectangle on the surface of
$O$ connecting two band rims whose top and bottom edges are
gridedges. In the degenerate case, a $z$-beam has height zero and
connects two rims along a section where they coincide. We say that
two bands $b_1$ and $b_2$ are {\em z-visible} if there exists a
$z$-beam connecting a gridedge of $b_1$ to a gridedge of $b_2$.
There can be many $z$-beams connecting two bands, so for each pair
of bands we select a representative $z$-beam of minimal (vertical)
height. Let $G$ be the graph that contains a node for each band of
$O$ and an arc for each pair of $z$-visible bands. It easily follows
from the connectedness of the surface of $O$ that $G$ is connected.
Let the unfolding tree $T_U$ be any spanning tree of $G$, with the
root selected arbitrarily from among all bands adjacent to $Y_0$.

Applying the 3x1 refinement partitions each front, back, top and
bottom face of $O$ into a $3 \times 1$ grid of faces. This
partitions the top and bottom edges of each $z$-beam into three
refined gridedges and divides the beam itself into three vertical
columns of refined gridfaces. For a band $B$ in $T_U$ with parent
$A$, let $e$ be the gridedge on $B$'s rim where the $z$-beam from
$A$ attaches. We define the {\em pivot point} $x_b$ to be the
$\frac{1}{3}$-point of $e$ (or, in circumstances to be explained
below, the $\frac{2}{3}$-point), and so it coincides with a point of
the $3 \times 1$-refined grid. The unfolding of $O$ will follow the
connecting vertical ray that extends from $x_b$ on $B$ to $A$. Note
that if $e$ belongs to both $A$ and $B$, then the ray connecting $A$
and $B$ degenerates to a point. To either side of a connecting ray
we have two \emph{connecting paths} of vertical faces, the
\emph{forward} and \emph{return} path. In
Fig.~\ref{fig:3x1.three.boxes}a, these connecting paths are the
shaded strips on the front face of $A$.

\subsection{Unfolding Bands into a Net}
\label{sec:3x1band}

\begin{figure}[htbp]
\centering
\includegraphics[width=0.9\linewidth]{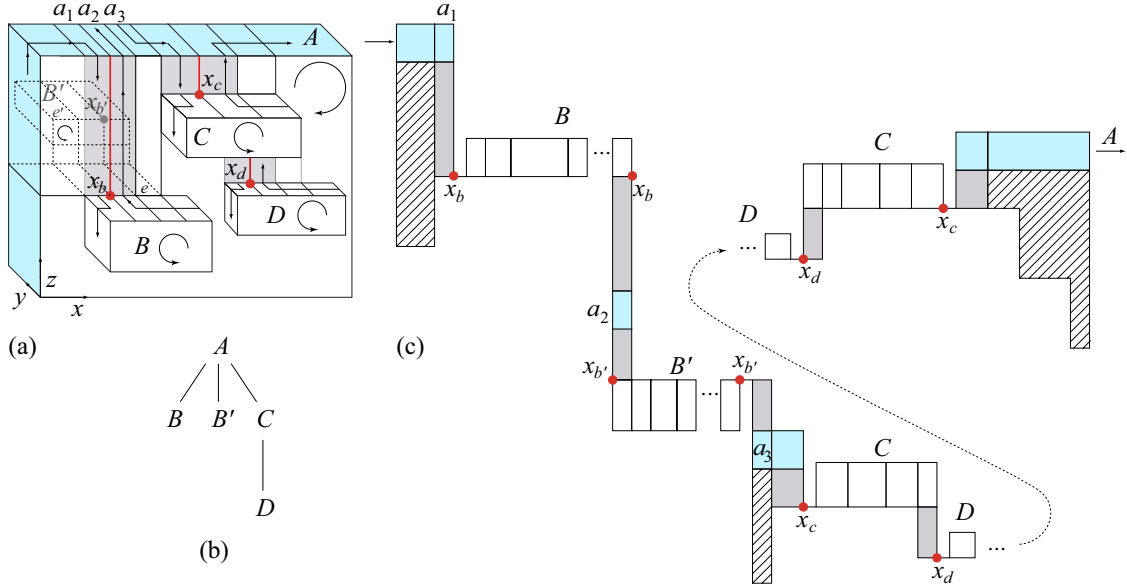}\\
\vspace{-0.08in} \caption{ (a)~Orthogonal polyhedron. (b)~Unfolding
tree $T_U$. (c)~Unfolding of bands and front (hachured) face pieces.
Vertex connection through the pivots points $x_b$, $x_{b'}$, $x_c$,
$x_d$ is shown exaggerated for clarity. }
\label{fig:3x1.three.boxes}
\end{figure}

Starting at a frontmost \emph{root band}, each band is unfolded as a
conceptual unit, but interrupted by the connecting rays incident to
it from its front and back faces. In Fig.~\ref{fig:3x1.three.boxes},
band $A$ is unfolded as a rectangle, but interrupted at the rays
connecting to (front children) $B$, $C$ and (back child) $B'$. At
each such ray the parent band unfolding is suspended, the unfolding
follows the forward connecting path to the child, the child band is
recursively unfolded, then the unfolding returns along the return
connecting path back to the parent, resuming the parent band
unfolding from the point it left off.

Fig.~\ref{fig:3x1.three.boxes} illustrates this unfolding algorithm.
The cw unfolding of $A$, laid out horizontal to the right, is
interrupted to traverse the forward path down to $B$, and $B$ is
then unfolded as a rectangle (composed of its contiguous faces). The
base $x_b$ of the connecting ray is called a \emph{pivot point}
because the ccw unfolding of $B$ is rotated $180^\circ$ ccw about
$x_b$ so that the unfolding of $B$ is also horizontal to the right.
It is only here that we use point-connections that render the
unfolding a vertex-unfolding. The unfolding of $B$ proceeds ccw back
to $x_b$, crosses over $A$ to unfold $B'$, then a cw rotation by
$180^\circ$ around the second image of pivot $x_b'$ orients the
return path to $A$ so that the unfolding of $A$ continues horizontal
to the right. Note that the unfolding of $C$ is itself interrupted
to unfold child $D$. Also note that there is edge overlap in the
unfolding at each of the pivot points.

The reason for the $3 \times 1$ refinement is that the upper edge
$e'$ of the back child band $B'$ has the same $(x,z)$-coordinates as
the upper edge $e$ of $B$ on the front face. In this case, the faces
of band $A$ induced by the connecting paths to $B$ would be
``overutilized'' if there were only two.  Let $a_1, a_2, a_3$ be the
three faces of $A$ induced by the $3 \times 1$ refinement of the
connecting path to $B$, as in Fig.~\ref{fig:3x1.three.boxes}. Then
the unfolding path winds around $A$ to $a_1$, follows the forward
connecting path to $B$, returns along the return connecting path to
$a_2$, crosses over $A$ and unfolds $B'$ on the back face, with the
return path now joining to $a_3$, at which point the unfolding of
$A$ resumes.  In this case, the pivot point $x_{b'}$ for $B'$ is the
$\frac{2}{3}$-point of $e'$. Other such conflicts are resolved
similarly. It is now easy to see that the
resulted net has the general form illustrated in
Fig.~\ref{fig:3x1.three.boxes}b:
\begin{enumerate}
\squeezelist
\item The faces of each band fall within a horizontal rectangle whose height is the band
width.
\item These band rectangles are joined by vertical connecting paths on either side,
connecting through pivot points.
\item The strip of the plane above and below each band face that is not incident to a
connecting path, is empty.
\item The net is therefore an orthogonal polygon monotone with respect to
the horizontal.
\end{enumerate}

\subsection{Attaching Front and Back Faces to the Net}
\label{sec:3x1front.back}

Finally, we ``hang'' front and back faces from the bands as follows.
The front face of each band $A$ is partitioned by imagining $A$ to
illuminate downward lightrays from the rim in the front face. The
pieces that are illuminated are then hung vertically downward from
the horizontal unfolding of the $A$ band. The portions unilluminated
will be attached to the obscuring bands.

In the example in Fig.~\ref{fig:3x1.three.boxes}, this illumination
model partitions the front face of $A$ into three pieces (the
striped pieces in Fig.~\ref{fig:3x1.three.boxes}b). These three
pieces are attached under $A$; the portions of the front face
obscured by $B$ but illuminated downward by $B$ are hung beneath the
unfolding of $B$ (not shown in the figure), and so on. Because the
vertical illumination model produces vertical strips, and because
the strips above and below the band unfoldings are empty, there is
always room to hang the partitioned front face. Thus, any orthogonal
polygon may be vertex-unfolded with a $3 \times 1$ refinement of the
vertex grid.

\medskip
\noindent Although we believe this algorithm can be improved to $2
\times 1$ refinement, the complications needed to achieve this are
similar to what is needed to avoid refinement entirely, so we
instead turn directly to $1 \times 1$ refinement.

\section{$(1 \times 1)$-Algorithm}

Although the $(1 \times 1)$-algorithm follows the same general
outline as the $(3 \times 1)$-Algorithm, there are significant
complications, which we outline before detailing. First, without the
refinement of $z$-beams into three strips to allow avoidance of
conflicts on opposite sides of a slab (e.g., $B$ and $B'$ in
Fig.~\ref{fig:3x1.three.boxes}a), we found it necessary to replace
the $z$-beams by a pair of $z$-rays that are in some sense the
boundary edges of a $z$-beam. Selecting two rays per band permits a
2-coloring algorithm (Theorem~\ref{thm:2-color}) to identify rays
that avoid conflicts. Generating the ray-pairs
(Sec.~\ref{sec:ray-pairs}) requires care to ensure that the band
graph $G_b$ is connected (Sec.~\ref{sec:Gb.connected}). This graph,
and the 2-coloring, lead to an unfolding tree $T_U$
(Sec.~\ref{sec:spanning.tree}). From here on, there are fewer
significant differences compared to the $(3 \times 1)$-Algorithm.
Without the luxury of refinement, there is more need to share
vertical paths on the vertical face of a slab
(Fig.~\ref{fig:return}).
Finally, the vertical connecting paths obscure the illumination of
some grid faces, which must be attached to the connecting paths. We
now present the details, in this order:


\vspace{0.1in} \hrule \hfill \vspace{-0.1in}
\begin{tabbing}
....\=...........\=........\kill
       \> 1. Select Pivot Points (Sec.~\ref{sec:pivots}) via \\
       \>      \>a. Ray-Pair Generation (Sec.~\ref{sec:ray-pairs}) \\
       \>      \>b. Ray Graph (Sec.~\ref{sec:ray.graph}) \\
       \> 2. Construct $T_U$ (Sec.~\ref{sec:spanning.tree}) \\
       \> 3. Select Connecting Paths (Sec.~\ref{sec:paths}) \\
       \> 4. Determine Unfolding Directions (Sec.~\ref{sec:unfdir}) \\
       \> 5. Recurse: \\
       \>      \> a. Unfold Bands into a Net (Sec.~\ref{sec:unfolding.bands}) \\
       \>      \> b. Attach Front and Back Faces to the Net (Sec.~\ref{sec:faces})
\end{tabbing}
\hrule \hfill

\subsection{Selecting Pivot Points} \label{sec:pivots}
The pivot $x_a$ for a band $A$ is the gridpoint of $A$ where the
unfolding of $A$ starts and ends. The $y$-edge of $A$ incident to
$x_a$ is the first edge of $A$ that is cut to unfold $A$.

Let $A$ be an arbitrary band delimited by planes $Y_i$ and
$Y_{i+1}$. Say that two gridpoints $u \in Y_i$ and $w \in Y_{i+1}$
are {\em in conflict} if the upward rays emerging from $u$ and $w$
hit the endpoints of the same $y$-edge of $A$; otherwise, $u$ and
$v$ are {\em conflict-free}. If $u$ lies either on a vertical edge,
or on a vertically extreme horizontal edge, then the ray at $u$
degenerates to $u$ itself.

Our goal is to select conflict-free pivots for all bands in $T_U$,
which will help us avoid later competition over the use of certain
faces in the unfolding, an issue that will become clear in
Sec.~\ref{sec:unfolding.bands}. Selecting these pivots is the most
delicate aspect of the $(1 \times 1)$-algorithm. Ultimately, we
represent pivoting conflicts in the form of a graph $G_r$
(Sec.~\ref{sec:ray.graph}), from which $T_U$ will be derived.

\subsubsection{Ray-Pair Generation}
\label{sec:ray-pairs} 
\begin{figure}[htbp]
\centering
\begin{tabular}{c@{\hspace{0.15\linewidth}}c}
\includegraphics[width=0.3\linewidth]{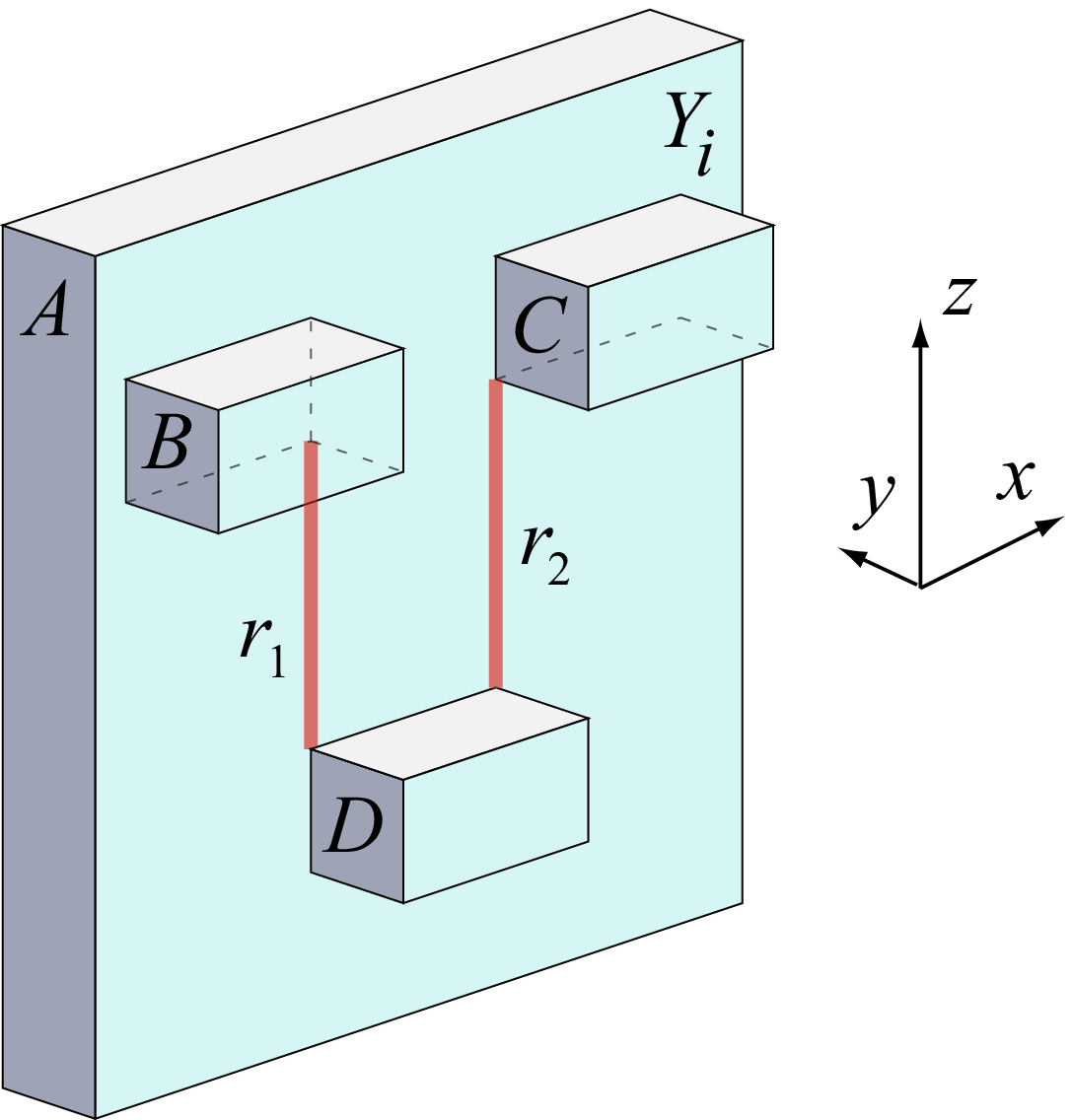} &
\includegraphics[width=0.34\linewidth]{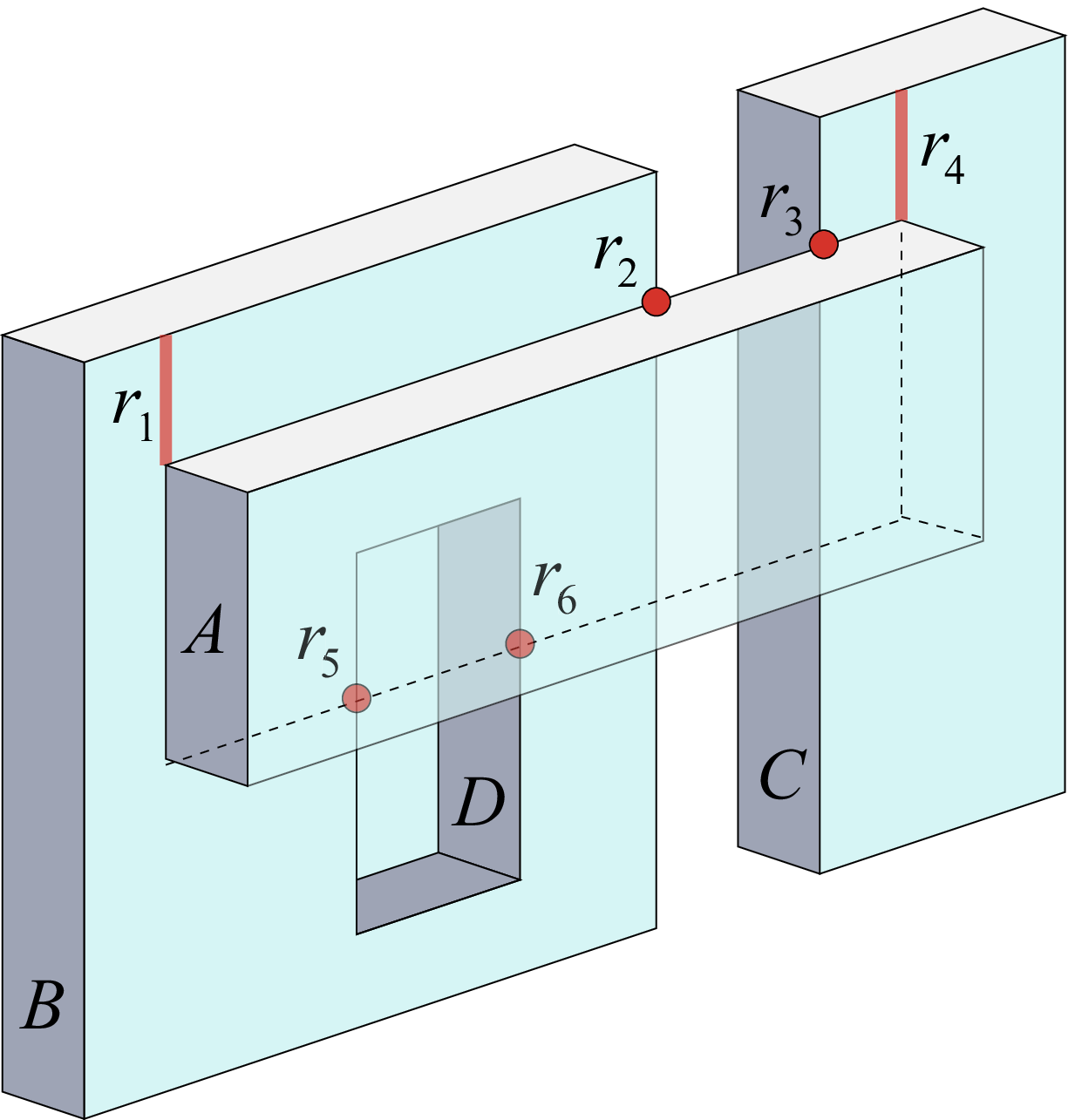} \\
(a) & (b)
\end{tabular}
\caption{(a) The ray pair $(r_1, r_2)$ connects band $D$ to two
different bands $B$ and $C$. (b) To ensure connectivity, three pairs
of rays must be issued for $A$: $(r_1, r_2)$, $(r_3, r_4)$ and
$(r_5, r_6)$.} \label{fig:face}
\end{figure}

In order to avoid pivoting conflicts, for each
band we will need two choices for its connecting ray. Thus the
algorithm generates the rays in pairs. Because there is no
refinement, the two rays originate at grid points on the same band,
but they may terminate on different bands. A simple example is shown
in Figure~\ref{fig:face}a, where the ray pair originating on band
$D$ hits two different bands, $B$ and $C$. This example also
suggests that one cannot consider ray pairs connecting pairs of
bands, as in the $(3 \times 1)$-algorithm (which would connect $D$
to $A$ in this example), but instead we focus on shooting pairs of
rays upward from strategic locations on the boundary of each band,
and then selecting a subset of these rays so that the conflicts can
be resolved and $T_U$ is connected. To ensure connectedness of all
bands, several ray-pairs must be issued upward from each band.
Figure~\ref{fig:face}b shows an example: no pair of rays can emanate
upward from the top of $B \cap P^-$ or $C \cap P^-$; one pair of
rays shoots upward from the top of each component of $A \cap P^-$:
$(r_1, r_2)$ connects $A$ to $B$ and $(r_3, r_4)$ connects $A$ to
$C$; finally, one pair of rays $(r_5, r_6)$ issues from the top of
$A \cap P^-$, which connects $A$ to $D$. So, overall, three pairs of
rays are generated for band $A$. We now turn to describing in detail
the method for generating ray-pairs.
%

\begin{figure}[htbp]
\centering
\begin{tabular}{c@{\hspace{0.1\linewidth}}c@{\hspace{0.1\linewidth}}c}
\includegraphics[width=0.24\linewidth]{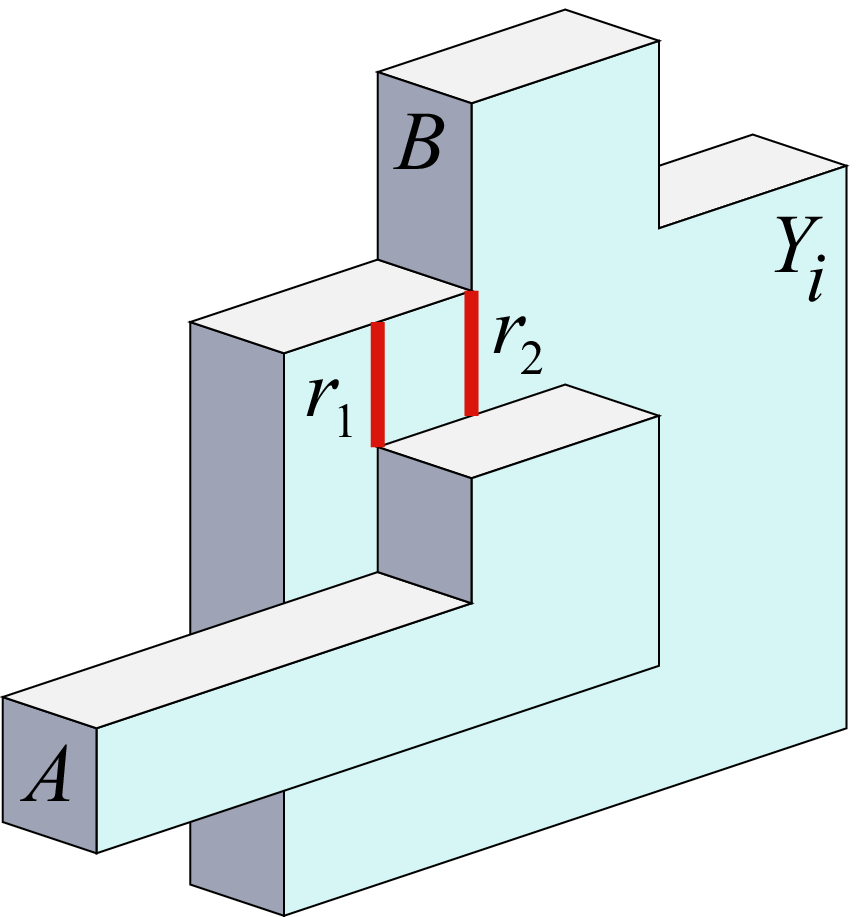} &
\includegraphics[width=0.25\linewidth]{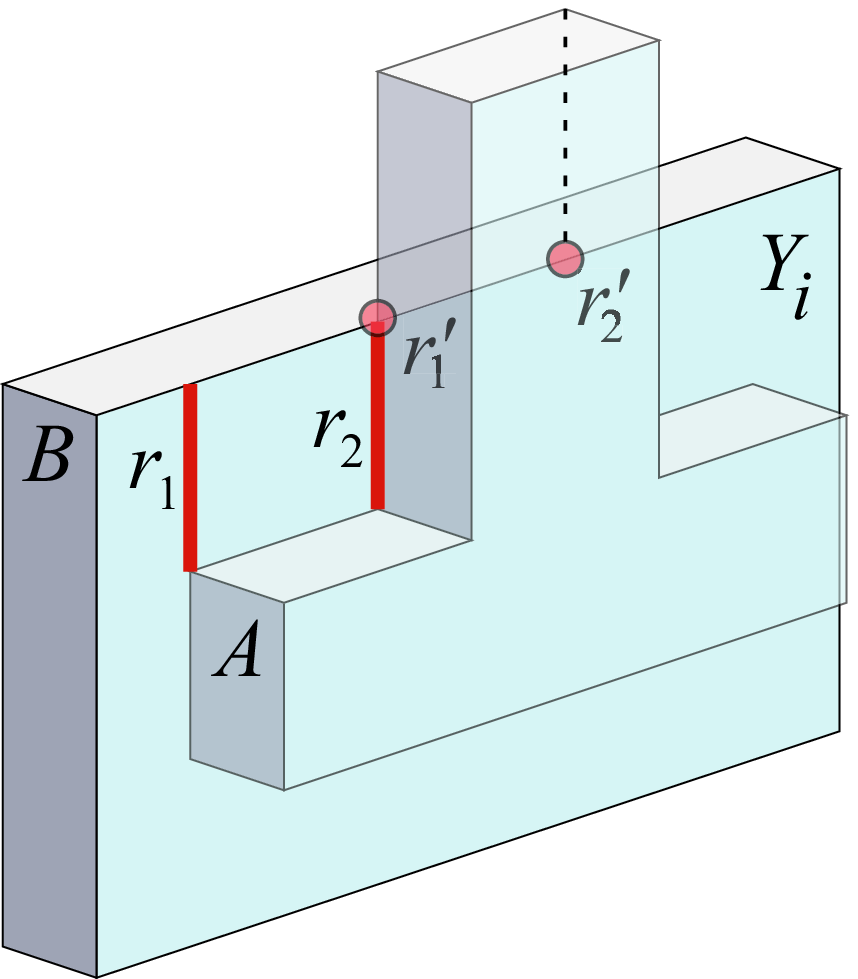} &
\includegraphics[width=0.21\linewidth]{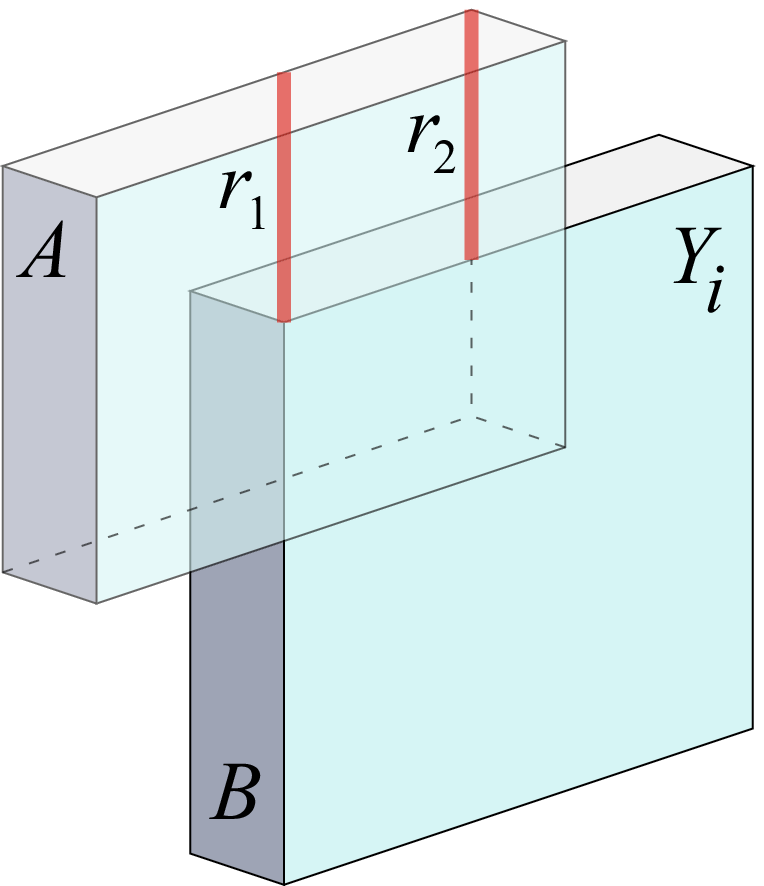} \\
(a) & (b) & (c)
\end{tabular}
\caption{Generating ray-pairs: (a)~$(r_1, r_2)$ for $A$; $S(B) =
\emptyset$. (b)~$(r_1, r_2)$ for $A$ (note that $r_2$ runs along
source band $A$); degenerate ray-pair $(r'_1, r'_2)$ for $B$.
(c)~$S(A) = \emptyset$; $(r_1, r_2)$ for $B$. } \label{fig:ray.pairs}
\end{figure}

Let band $A$ intersect plane
$Y_i$. The algorithm is a for-loop over all $A$. Let $A_1, A_2,
\ldots, A_m$ be the components of $A$, defined as follows. Take all
the maximal components of $A \cap P^+$ that contain an $x$-gridedge,
and union with all the maximal components of $A \cap P^-$ that
contain an $x$-gridedge. We define $S(A_j)$ as the set of all
potential rays shooting upward from $A_j$. More precisely, $S(A_j)$
consists of the set of all segments $s=(a,b)$, with $a \in A_j$,
such that
\begin{enumerate}
\squeezelist
\item Either $s$ is a point, with $b=a$, or
$s \subset P \subset Y_i$ is vertical (parallel to $z$), with $a$
below $b$.
\item $b \in B$ for some band $B \neq A$.
\item The open segment $s \setminus \{a,b\}$ may contain points of $A$
(see $r_2$ in Fig.~\ref{fig:ray.pairs}b), but no other band points.
\end{enumerate}
%
For each band $A$, for each component $A_j \subseteq A$, if $S(A_j)$
is nonempty, we select one ray pair $(r_1, r_2)$, such that (i)
$r_1$ is the leftmost segment in $S(A_j)$ that is incident to a
highest $x$-gridedge in $A_j$, and (ii) $r_2$ is the segment one
$x$-gridedge to the right of $r_1$. Fig.~\ref{fig:ray.pairs} shows a
few examples.
As mentioned above, several ray pairs could be generated for any one
band, and indeed several pairs connecting two bands.

Let $G_b$ be the \emph{band graph} whose nodes are bands. Two bands
are connected by an arc in $G_b$ if the ray-pair algorithm generates
a ray connecting them.  We call a collection of bands in $G_b$
\emph{ray-connected} if they are in the same connected component of
$G_b$. We establish that $G_b$ is a connected graph, i.e., all bands
are ray-connected to one another, even if only one ray per pair is
employed:

\begin{lemma}
$G_b$ is connected. Furthermore, the subgraph of $G_b$ induced by
exactly one ray per ray-pair (arbitrarily selected) is connected.
\label{lem:Gb.connected0}
\end{lemma}

\medskip
\noindent Whereas the connectedness of bands by $z$-beams in the $(3
\times 1)$-algorithm is straightforward, the complex possible
relationships between bands makes connectedness via rays more
subtle. We relegate the proof to the Appendix
(Sec.~\ref{sec:Gb.connected}) in order to not interrupt the main
flow of the algorithm.

The over-generation of ray-pairs noted above is designed to ensure
connectedness. Eventually many rays will be discarded by the time
$T_U$ is constructed in Sec.~\ref{sec:spanning.tree}.

\subsubsection{Ray Graph $G_r$} \label{sec:ray.graph}
\begin{figure}[htbp]
\centering
\begin{tabular}{c@{\hspace{0.1\linewidth}}c}
\includegraphics[width=0.4\linewidth]{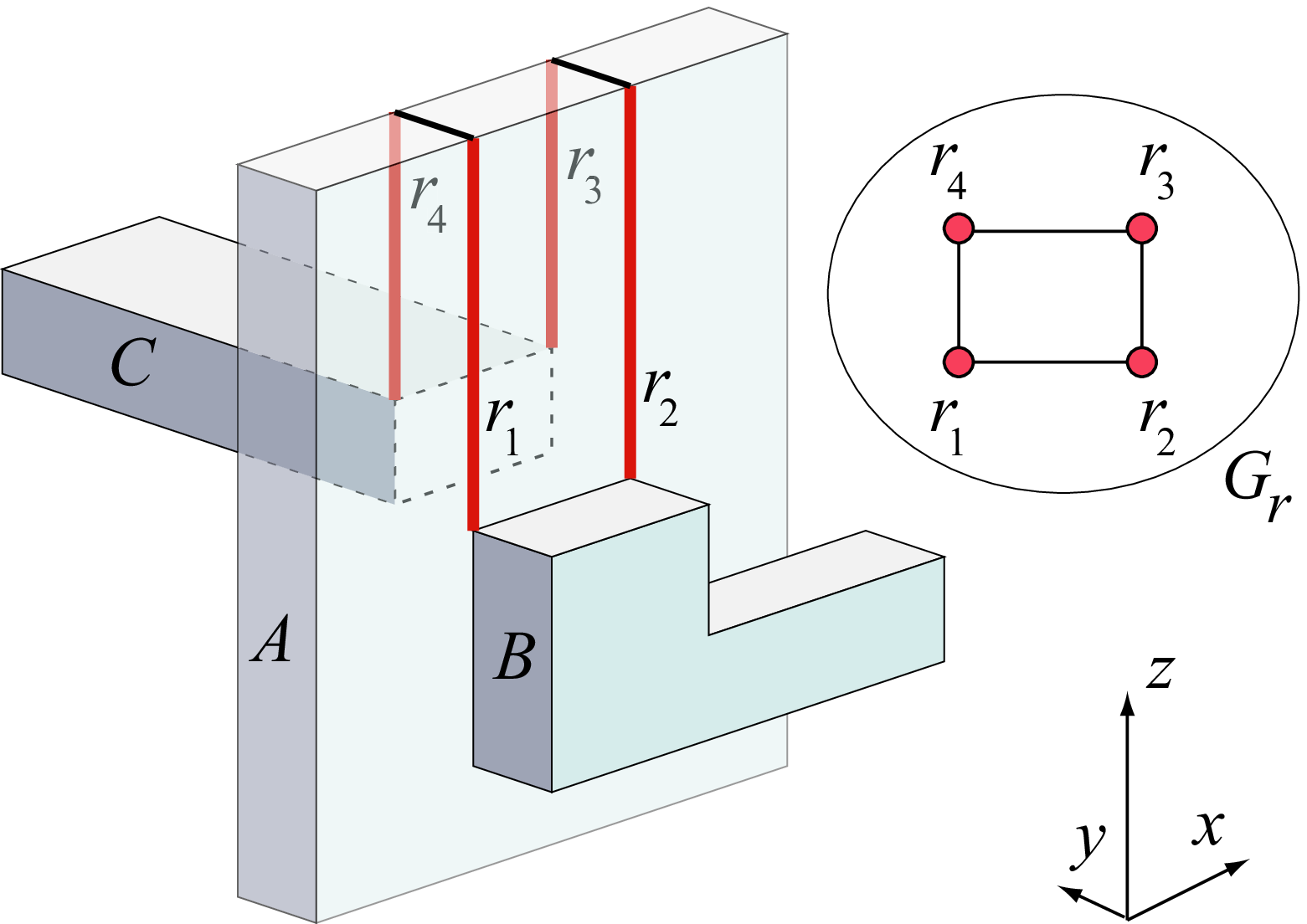} &
\includegraphics[width=0.4\linewidth]{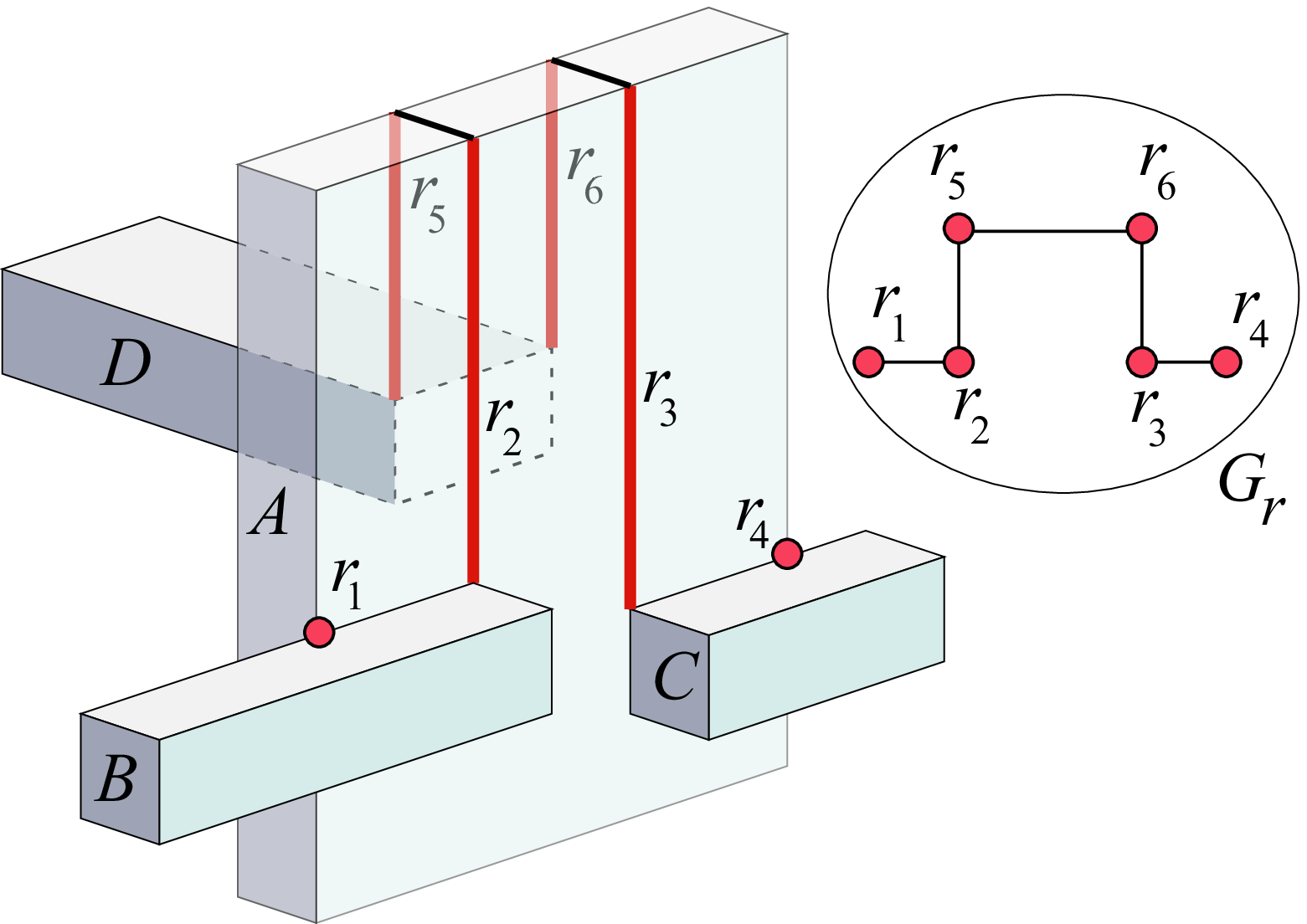} \\
(a) & (b)
\end{tabular}
\caption{ Building $G_r$. (a) $G_r$ is a $4$-cycle; $\{r_1,r_2\}$
and $\{r_3, r_4\}$ are $x$-arcs; any other arc is a $y$-arc. (b)
$G_r$ is a path; $\{r_2,r_5\}$ and $\{r_3, r_6\}$ are $y$-arcs; any
other arc is an $x$-arc.} \label{fig:Gr.path}
\end{figure}

One pair of rays per pair of bands suffices to ensure that all bands
are ray-connected. If multiple pairs of rays exist for a pair of
bands, pick one pair arbitrarily and discard the rest. Then define a ray
graph $G_r$ as follows.  The nodes of $G_r$ are vertical rays in a
plane $Y_i$, perhaps degenerating to points, connecting gridpoints
between two bands that both intersect $Y_i$. The arcs of $G_r$ each
records a potential pivoting conflict, and are of two varieties:

\begin{enumerate}
\squeezelist
\item[(i)] The nodes for the two rays issuing from the top of one band $B$
are adjacent in $G_r$.  Call such arcs \emph{$x$-arcs};
geometrically they can be viewed as parallel to the $x$-axis.

\item[(ii)] The nodes for two rays incident to opposite sides of
the rim of a band $A$, connected by a $y$-segment on the band, are
adjacent in $G_r$.  Call such arcs \emph{$y$-arcs}; geometrically
they can be viewed as parallel to the $y$-axis.

\end{enumerate}
Fig.~\ref{fig:Gr.path} shows two simple examples of $G_r$ involving
nodes on opposite sides of one band $A$.
%
Before proceeding, we list the consequences of the two types of arcs
in $G_r$. Assuming that we can $2$-color $G_r$ $\{$red, blue$\}$,
and we select the base of (say) the red rays as pivots, then: (i)
exactly one pivot is selected for each band, and (ii) no two pivot
rays are in conflict across a band. So our goal now is to show that
$G_r$ is $2$-colorable. Because a graph is $2$-colorable if and only
if it is bipartite, and a graph is bipartite if and only if every
cycle is of even length, we aim to prove that every cycle in $G_r$
is of even length. We start by listing a few relevant properties of
$G_r$:
\begin{enumerate}
\squeezelist
\item
Every node $r \in G_r$ has exactly one incident $x$-arc. The rays
are generated in pairs, and the pairs are connected by an $x$-arc.
As no such ray is shared between two bands, at most one $x$-arc is
incident to any $r$.

\item Nodes have at most degree $3$, with the following structure:
degree-$1$ nodes have  an incident $x$-arc; degree-$2$ nodes have
both an incident $x$- and $y$-arc; and degree-$3$ nodes have an
incident $x$-arc and two incident $y$-arcs.


\item Each $x$-arc spans exactly one pair of adjacent $y$-gridlines,
and each $y$-arc spans exactly one band rim-to-rim. The former is by
the definition of ray pairs, which issue from adjacent gridpoints,
and the latter follows from the grid partitioning of the object into
bands.
\end{enumerate}
%
\begin{figure}[htbp]
\centering
\includegraphics[width=0.98\linewidth]{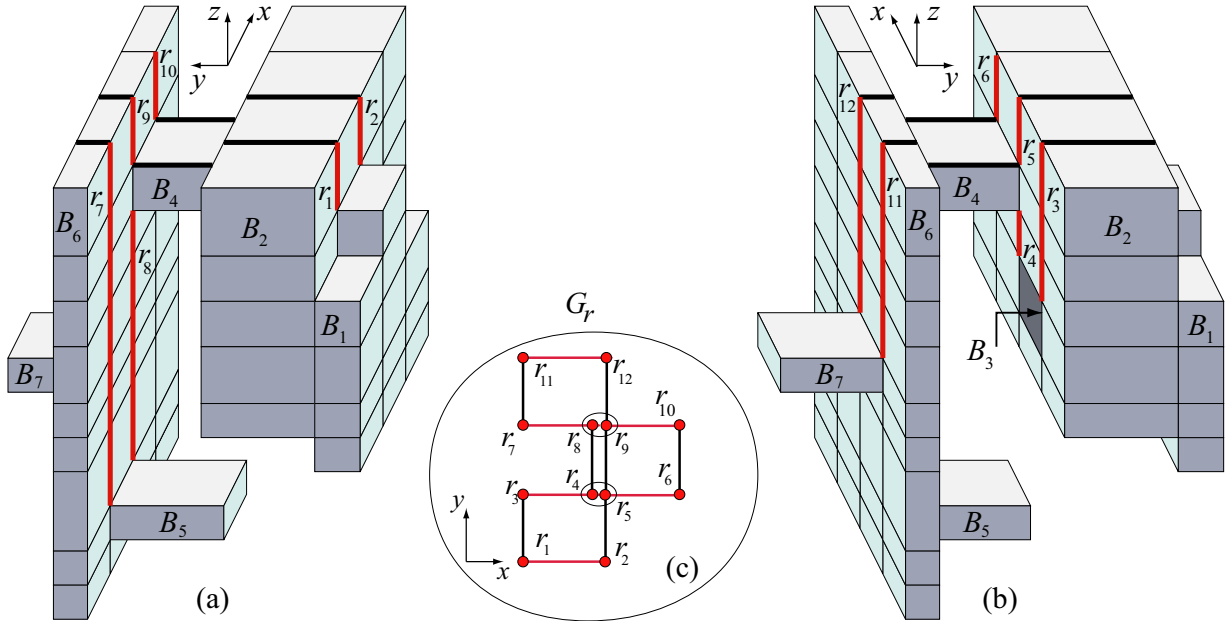}
\caption{ (a,b) Two side views of an object;  $z$-rays and $y$-arcs
are marked with thick lines. (c) $G_r$ coordinatized into $xy$-plane
$\Pi$; $(r_5, r_6, r_{10}, r_9)$ is a $4$-cycle;
$(r_1,r_3,r_4,r_8,r_7,r_{11},r_{12},r_9,r_5,r_2)$ is a $10$-cycle.}
\label{fig:dent.10-cycle}
\end{figure}
%
Our next step requires embedding $G_r$ in an $xy$-plane $\Pi$.
Toward that end, we coordinatize the nodes and arcs of $G_r$ as
follows. A node $r \in G_r$ is a $z$-ray, and is assigned the
$(x,y)$ coordinates of the ray.  Note that this means collinear rays
get mapped to the same point; however we treat them as distinct. The
$x$-arcs are then parallel to the $x$-axis, and the $y$-arcs are
parallel to the $y$-axis. In essence, this coordinatization is a
view from $z=+\infty$.

Fig.~\ref{fig:dent.10-cycle} shows a more complex example
illustrating this viewpoint. The object is composed of $7$ bands
$B_i$, one of which ($B_3$) is a dent. There are $12$ ray nodes, two
pairs of which lie on the same $z$-vertical line, namely $(r_4,r_5)$
and $(r_8,r_9)$. Note that there are $y$-arcs crossing both the top
of and the bottom\footnote{
    A dent is included in this example precisely to
    introduce such a bottom $y$-arc into $G_r$.
} of $B_4$. The graph $G_r$ has a $4$-cycle and a $10$-cycle, both
detailed in the caption (as well as a $12$-cycle not detailed).

\begin{lemma}
Every cycle in $G_r$ is of even length.
\end{lemma}
\begin{proof}
Let $C$ be a cycle in $G_r$.  The coordinatization described above
maps $C$ to a (perhaps self-crossing) closed path in the $xy$-plane
$\Pi$, a path which may visit the same $(x,y)$ point more than once,
and/or traverse the same edge in $\Pi$  more than once. Any such
closed path on a grid must have even length, for the following
reason.

First, by Property~(3) above, each edge of the path in $\Pi$
connects adjacent grid lines: an edge never ``jumps over'' one or
more grid lines. Second, any such closed lattice path changes parity
with each step, in the following sense.
Number the $x$- and $y$-gridlines with integers $0,1,2,\ldots$ left
to right and bottom to top respectively. Define the parity of a
gridpoint of $\Pi$ to be the sum of its $x$- and $y$-gridline
coordinates, mod~$2$. Then each step of the path, necessarily in one
of the four compass directions, changes parity, as it changes only
one of $x$ or $y$. Returning to the start point to close the path
must return to the starting coordinates, and so to the same parity.
Thus, there must be an even number of parity changes along any
closed path. Therefore, $C$ has an even number of edges.
\end{proof}

\medskip
\noindent We have now established this:
\begin{theorem}
$G_r$ is $2$-colorable. \label{thm:2-color}
\end{theorem}

\noindent Note that nowhere in the above proof do we assume genus
zero, so this theorem holds for polyhedra of arbitrary genus.

\paragraph{Band pivoting.}
By Theorem~\ref{thm:2-color}, we can $2$-color the nodes of $G_r$
\{red,blue\}. We choose all red ray-nodes of $G_r$ to be pivoting
rays, in that their base points become pivot points.  As remarked
before, this selection guarantees that each band is pivoted, and no
two pivots are in conflict.

\subsection{Unfolding Tree $T_U$} \label{sec:spanning.tree}The next
task is to define a band spanning tree $T_U$, based on the band
graph $G_b$. Define $G'_b$, to retain the just the arcs of $G_b$
corresponding to the red ray nodes (in the above $2$-coloring) in
$G_r$. This maintains the connectivity of
Lemma~\ref{lem:Gb.connected0}.
Then take $T_U$ to be any
spanning tree of $G'_b$ rooted at a frontmost band.

With $T_U$ finally in hand, the remainder of the $(1 \times
1)$-algorithm follows the overall structure of the $3 \times 1$
algorithm, with variations as mentioned before, as detailed below.

\subsubsection{Selecting Connecting Paths}
\label{sec:paths} Having established a pivot point for each band, we
are now ready to define the {\em forward} and {\em return}
connecting paths for a child band in $T_U$. Let $B$ be an arbitrary
child of a band $A$. If $B$ intersects $A$, both forward and return
connection paths for $B$ reduce to the pivot point $x_b$ (e.g., $u$
in Fig.~\ref{fig:overhang}). If $B$ does not intersect $A$, then a
ray $r$ connects $x_b$ to $A$ (Figs.~\ref{fig:box-in-box1}a
and~\ref{fig:box-aligned-box}a). The connecting paths are the two
vertical paths separated by $r$ comprised of the gridfaces sharing
an edge with $r$ (paths $a_1$ and $a_2$ in
Figs.~\ref{fig:box-in-box1}a and~\ref{fig:box-aligned-box}a). The
path first encountered in the unfolding of $A$ is used as a forward
connecting path; the other path is used as a return connecting path.

\subsubsection{Determining Unfolding Directions}
\label{sec:unfdir} A top-down traversal of $T_U$ assigns an
unfolding direction to each band in $T_U$ as follows. The root band
in $T_U$ may unfold either cw or ccw, but for definiteness we set
the unfolding direction to cw. Let $B$ be the band in $T_U$
currently visited and let $A$ be the parent of $B$. If the upward
ray $r$ incident to $x_b$ connects $B$ to a bottom gridpoint of $A$,
and if $A$ unfolds cw(ccw), then $B$ unfolds cw(ccw). Otherwise, $r$
connects $B$ to a top or a side (for degenerate rays) gridpoint of
$A$; in this case, if $A$ unfolds cw(ccw), then $B$ unfolds ccw(cw).
In other words, $A$ and $B$ unfold in a same direction if $B$
``hangs below'' $A$, and in opposite direction otherwise.

\subsection{Unfolding Bands into a Net}
\label{sec:unfolding.bands} Let $A$ be a band to unfold, initially
the root band. The unfolding of $A$ starts at $x_a$ and proceeds in
the unfolding direction (cw or ccw) of $A$. Henceforth we assume
w.l.o.g. that the unfolding of $A$ proceeds cw (w.r.t. a viewpoint
at $y = -\infty$); the ccw unfolding of $A$ is a vertical reflection
of the cw unfolding of $A$. In the following we describe our method
to unfold every child $B$ of $A$ recursively, which falls naturally
into several cases.

\begin{figure}[htbp]
\centering \centering
\begin{tabular}{c@{\hspace{0.1\linewidth}}c}
\includegraphics[width=0.43\linewidth]{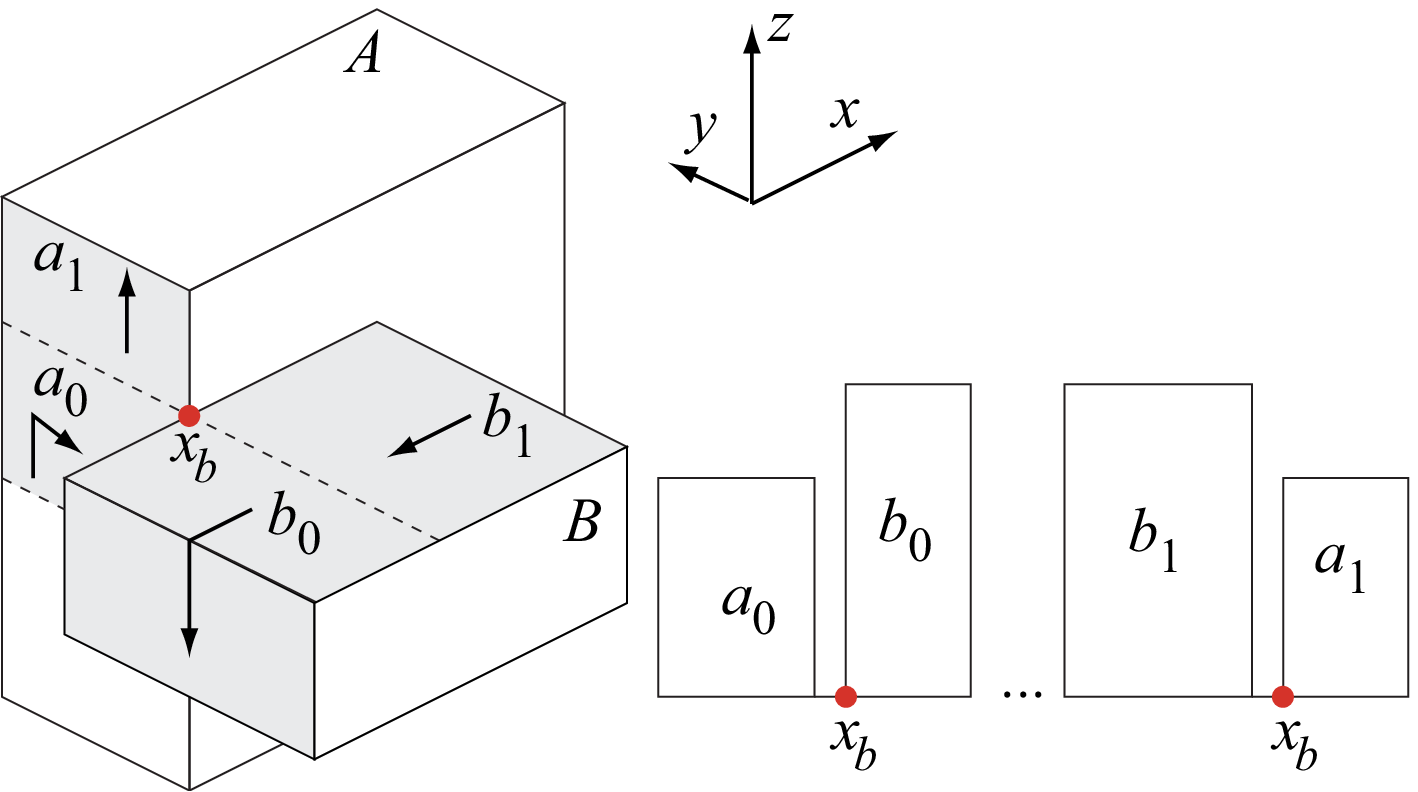} &
\includegraphics[width=0.42\linewidth]{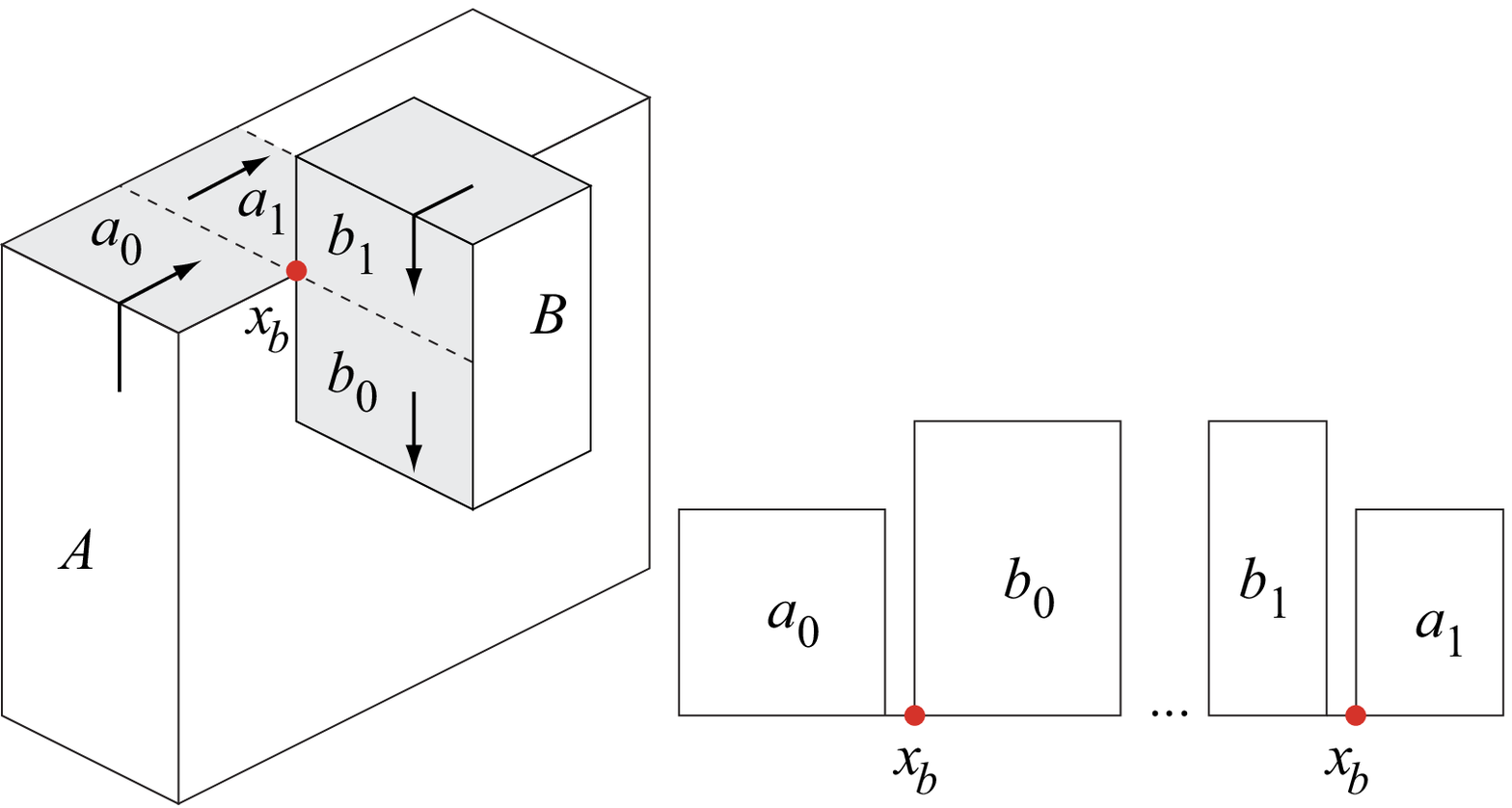}
\end{tabular}
\caption{Unfolding $B$ when the ray connecting $B$ to $A$
degenerates to $x_b$.} \label{fig:overhang}
\end{figure}

\paragraph{Case 1:} Pivot $x_b \in A \cap B$.
Then, whenever the unfolding of $A$ reaches $x_b$, we unfold $B$ as
in Fig.~\ref{fig:overhang}. The unfolding uses the two band faces of
$A$ incident to $x_b$ ($a_0$ and $a_1$ in Fig.~\ref{fig:overhang}).
The gridface $b_0$ of $B$ ccw of $x_b$ gets rotated around $x_b$ so
that the ccw unfolding of $B$ extends horizontally to the right. The
unfolding of $B$ proceeds ccw back to $x_b$, then the face $a_1$
incident to $x_b$ gets oriented about $x_b$ so that the unfolding of
$A$ continues horizontal to the right.

Note that, because the pivots of any two children of $A$ are
conflict-free, there is no competition over the use of $a_0$ and
$a_1$ in the unfolding. Note also that the unfolding path does not
self-cross. For example, the cyclic order of the faces incident to
$u$ in Fig.~\ref{fig:overhang}a is $(a_0,
A_{front},b_0,b_1,B_{back},a_1)$, and the unfolding path follows
$(a_0,b_0,\ldots,b_1,a_1)$.

\begin{figure}[htbp]
\centering
\begin{tabular}{c@{\hspace{0.2\linewidth}}c}
\includegraphics[width=0.28\linewidth]{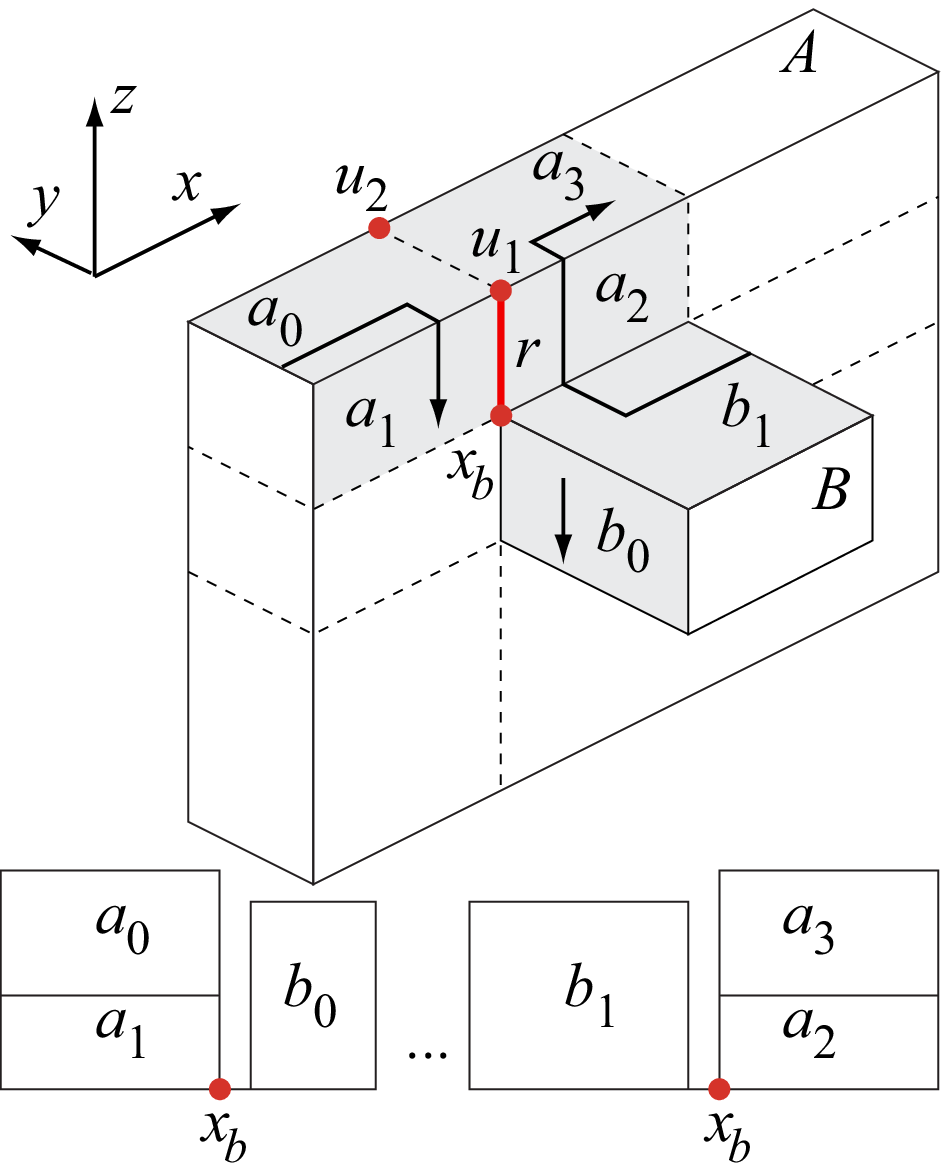} &
\includegraphics[width=0.32\linewidth]{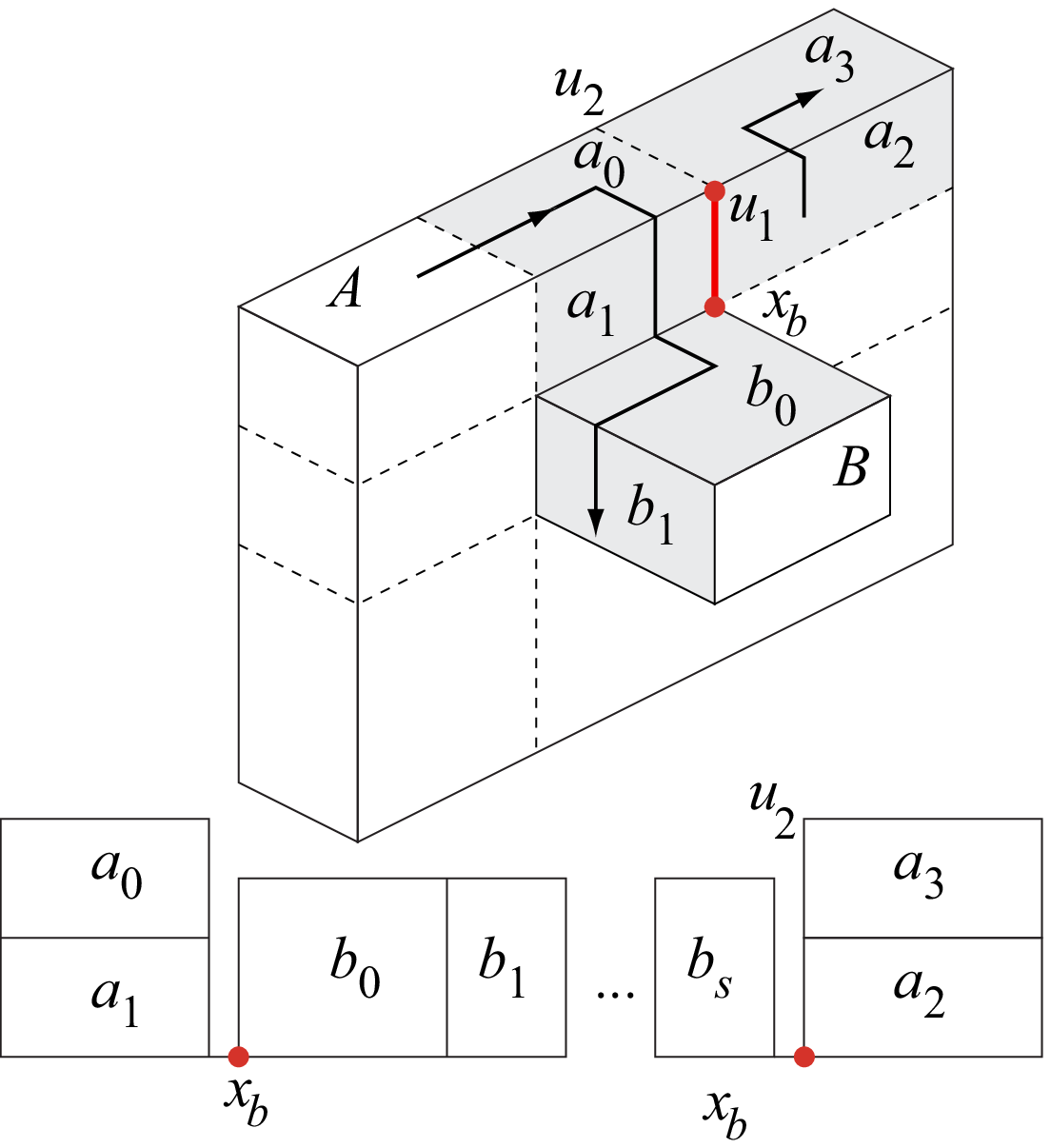} \\
(a) & (b)
\end{tabular}
\caption{Unfolding $B$: $u_1$ is not a corner vertex of $A$ (a)
$x_b$ incident to a left face of $B$ (b) $x_b$ incident to a top
face of $b$.} \label{fig:box-in-box1}
\end{figure}

\paragraph{Case 2:} Pivot $x_b \not\in A \cap B$ and
the (forward, return) connecting paths for $B$ do not overlap other
connecting paths (except at their boundaries);
we will later see that this may happen.
Let us settle some notation first 
(cf.~Fig~\ref{fig:box-in-box1}a):
~$r$ is the ray connecting $B$ to $A$; $a_1$ and $a_2$ are forward
and return connecting paths for $B$ (one to either side of $r$);
$u_1$ is the endpoint of $r$ that lies on $A$; and $u_2$ is the
other endpoint of the $y$-edge of $A$ incident to $u_1$.
%
We discuss three situations:

\paragraph{Case 2a:} $u_1$ is neither a reflex corner nor a bottom
corner of $A$. In this case, whenever the unfolding of $A$ reaches
$a_1$, the unfolding of $B$ proceeds as in
Fig.~\ref{fig:box-in-box1}a or Fig.~\ref{fig:box-in-box1}b,
depending on whether $x_b$ touches a left face of $B$ or not. In
either case, if $b_0$ is the face of $B$ extending ccw left of
$x_b$, rotate $b_0$ so that the unfolding of $B$ extends horizontal
to the right, recursively unfold $B$, then rotate the return path
$a_2$ about $x_b$ so that the unfolding of $A$ continues horizontal
to the right.

\begin{figure}[htbp]
\centering
\begin{tabular}{c@{\hspace{0.2\linewidth}}c}
\includegraphics[width=0.33\linewidth]{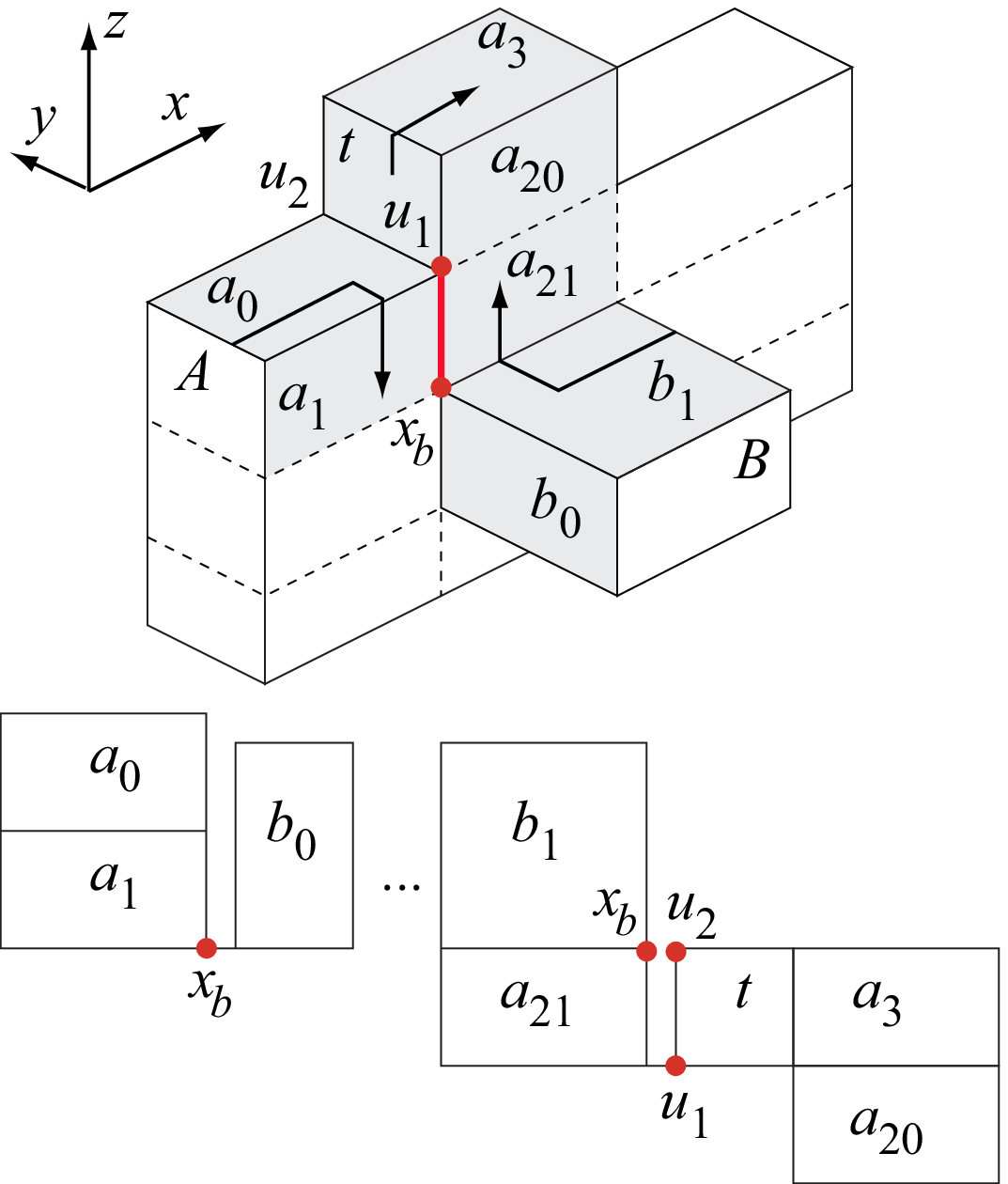} &
\includegraphics[width=0.27\linewidth]{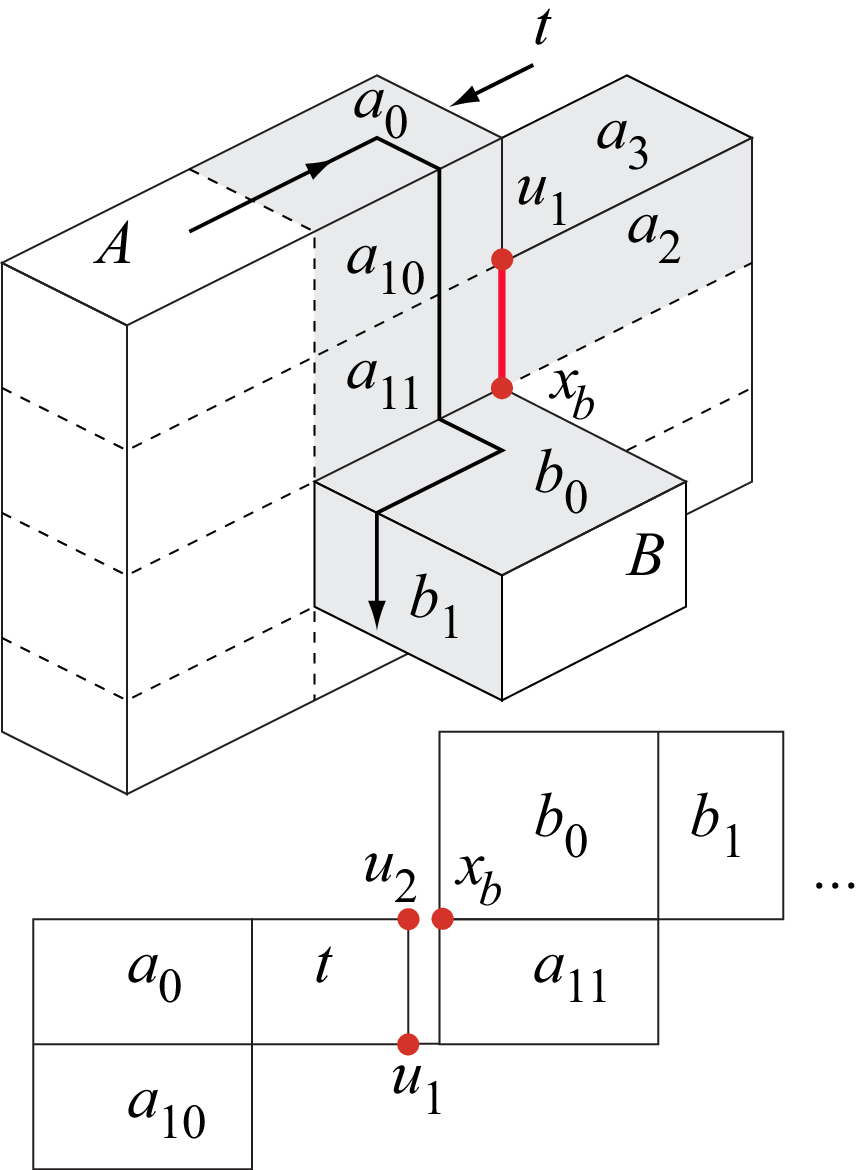} \\
(a)  & (b)
\end{tabular}
\caption{Unfolding $B$: $u_1$ is a corner vertex of $A$. (a) $t$ is
a left strip (b) $t$ is a right strip.} \label{fig:box-in-box2}
\end{figure}

\paragraph{Case 2b:} $u_1$ is a reflex corner of $A$. In this case,
the unfolding of $B$ proceeds as in Fig.~\ref{fig:box-in-box2}(a,
b). It is the existence of the vertical strip incident to $u_1$
(marked $t$ in Fig.~\ref{fig:box-in-box2}) that makes handling this
case different from Case 2a above. Note however that the existence
of $t$ implies the existence of at least two gridfaces on either the
return path or the forward path for $B$, depending on whether $t$ is
a left (Fig.~\ref{fig:box-in-box2}a) or a right
(Fig.~\ref{fig:box-in-box2}b) strip of faces. In the former case the
unfolding starts as in Case 2a (Fig.~\ref{fig:box-in-box2}a), and
once the unfolding of $B$ returns to $x_b$, it continues along the
return path up to $u_1$, then unfolds $t$ and orients it about $u_1$
in such a way that the unfolding of $A$ continues horizontal to the
right. The portion of the return path that extends above $u_1$
($a_{20}$ in Fig.~\ref{fig:box-in-box2}a) gets attached below the
adjacent top face of $A$ ($a_3$ in Fig.~\ref{fig:box-in-box2}a).
%
%

If $t$ is a strip of right faces, then $t$ gets unfolded before
descending along the forward path down to $B$, as in
Fig.~\ref{fig:box-in-box2}b (note the vertical symmetry with the
unfolding in Fig.~\ref{fig:box-in-box2}a); the unfolding of $B$ then
proceeds as in Case 2a (Fig.~\ref{fig:box-in-box1}b).

\begin{figure}[htbp]
\centering
\begin{tabular}{c@{\hspace{0.06\linewidth}}c}
\includegraphics[width=0.38\linewidth]{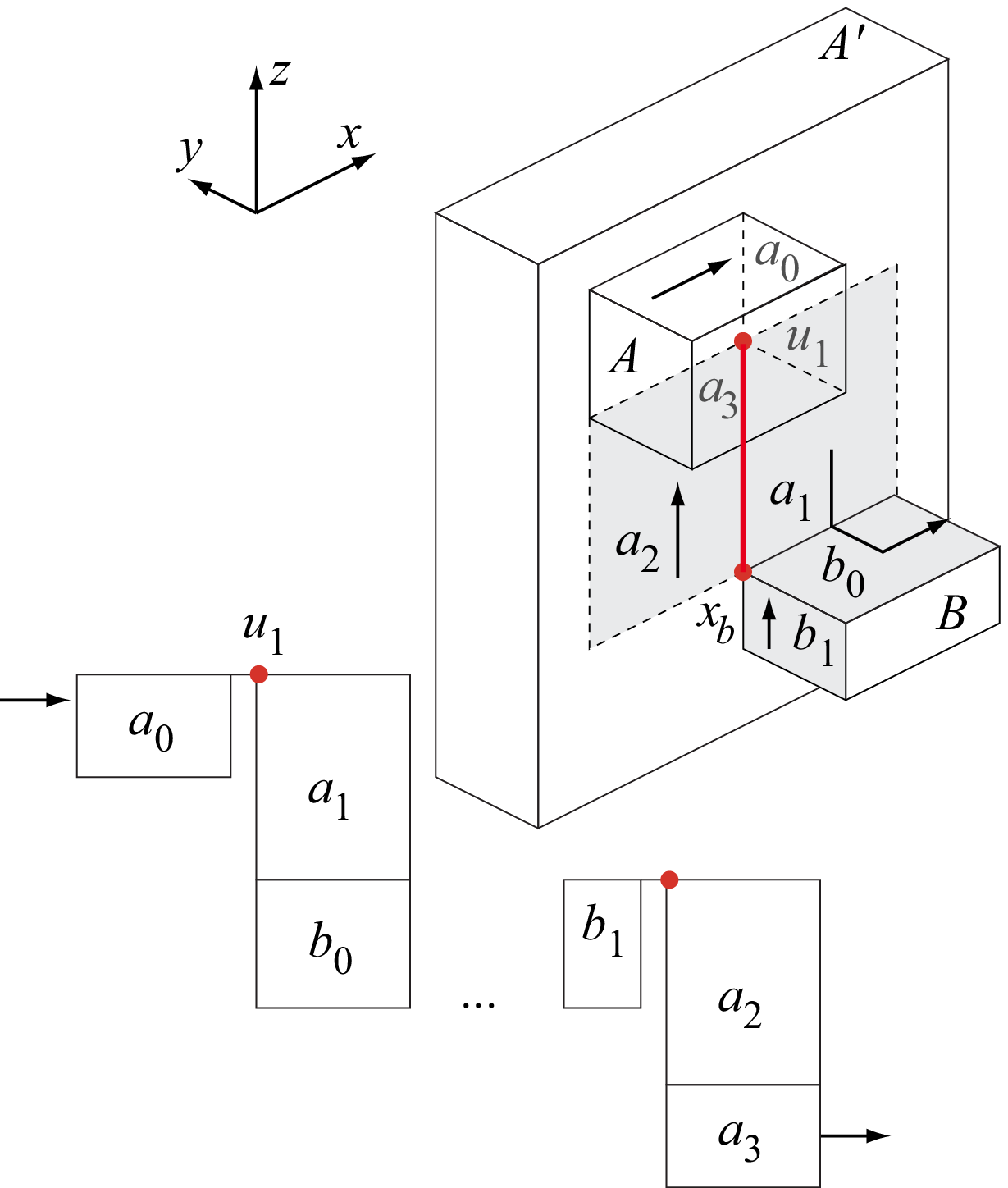}&
\includegraphics[width=0.44\linewidth]{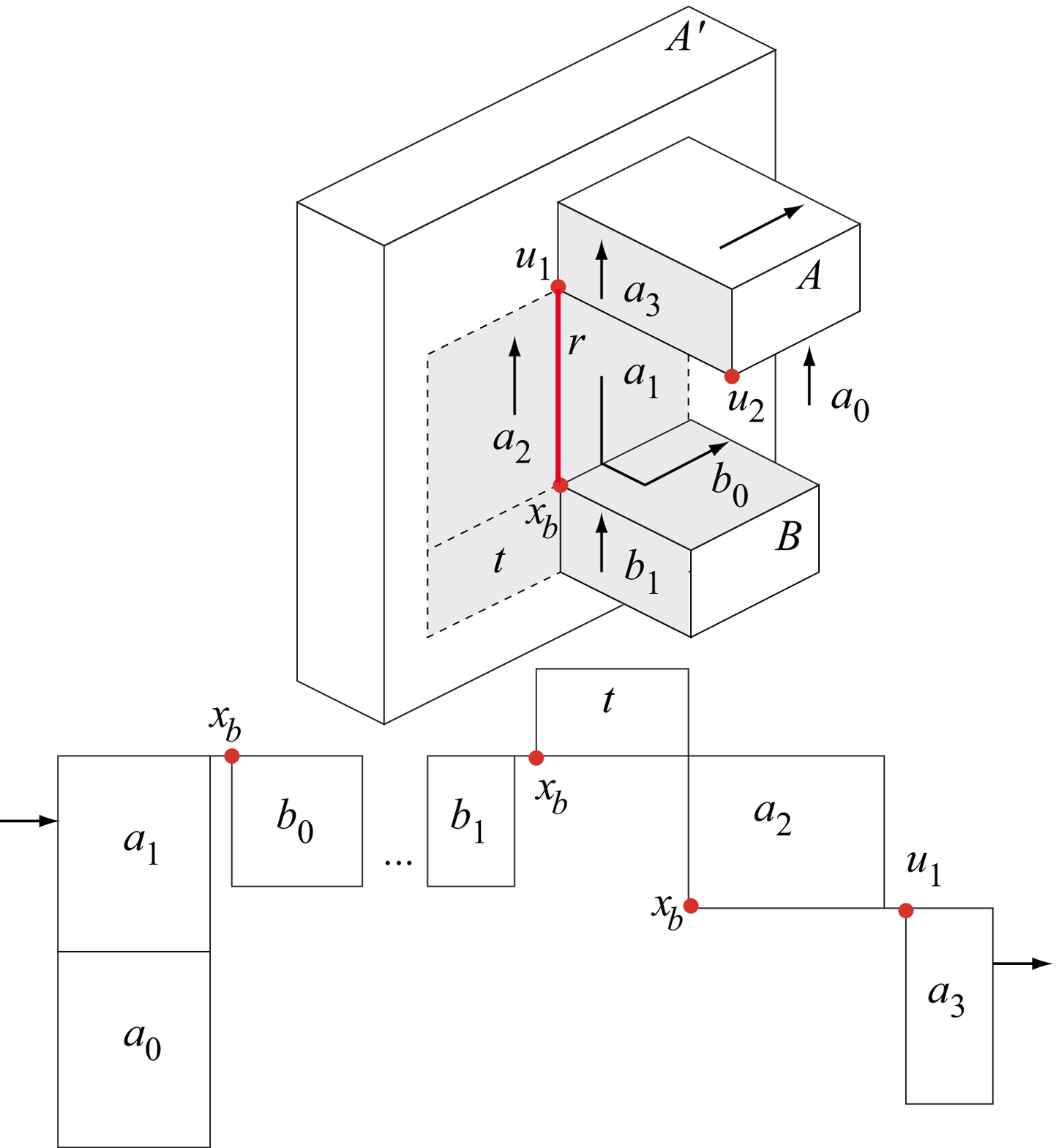}
\\
(a) & (b)
\end{tabular}
\caption{Unfolding $B$: $u_1$ is a bottom corner of $A$ (a)
rightmost, and (b) leftmost face of $A$ vertically aligned with
leftmost face of $B$.} \label{fig:box-aligned-box}
\end{figure}

\paragraph{Case 2c:} $u_1$ is a bottom corner of $A$. In this case,
the unfolding proceeds as in Fig.~\ref{fig:box-aligned-box}a or
Fig.~\ref{fig:box-aligned-box}b, depending on whether $u_1$ is a
right or a left bottom corner of $A$. The unfolding illustrated in
Fig.~\ref{fig:box-aligned-box}a follows the familiar unfolding
pattern: orient the face of $B$ ccw left of $x_b$ so that the
unfolding of $B$ extends to the right; once the unfolding of $B$
returns to $x_b$, follow the return path back to $A$ and unfold the
face of $A$ cw to the right of $u_1$ ($a_3$ in
Fig.~\ref{fig:box-aligned-box}a) so that the unfolding of $A$
continues horizontal to the right.
%
%
A similar pattern applies to the case illustrated in
Fig.~\ref{fig:box-aligned-box}b, with one subtle difference meant to
aid in unfolding front and back faces (discussed in
Sec.~\ref{sec:faces}): in unfolding bands, we aim at maintaining the
vertical position of the (forward, return) connecting paths in the
unfolding, so that vertical strips hanging below these connecting
paths could also hang vertically in the unfolding. More on this in
Sec.~\ref{sec:faces}. Observe that $a_1$  and $a_2$ from
Fig.~\ref{fig:box-aligned-box}a hang downward in the unfolding.
However, if $a_2$ were to maintain its vertical position in the
unfolding from Fig.~\ref{fig:box-aligned-box}b, it would not be
possible to orient $a_3$ around $u_1$ so as to continue unfolding
$A$ horizontal to the right of $a_2$. This is the reason for
employing the face marked $t$ in the unfolding, so that vertical
sides of $t$ remain vertical in the unfolding, and any face strip
hanging below $t$ could be attached to $t$ vertically in the
unfolding.

We note that Fig.~\ref{fig:box-aligned-box} illustrates only the
situation in which $x_b$ is incident to a left face of $B$, but it
should not be difficult to observe that an exact same idea applies
to any top pivot of $B$; the pivot position only affects the start
and end unfolding position of $B$, and everything else remains the
same.

\paragraph{Case 3:} Pivot $x_b \not\in A \cap B$ and a connecting
path for $B$ overlaps a connecting path for another descendant $C$
of $A$. This case is slightly more complex, because it involves
conflicts over the use of the connecting paths for $B$. The
following three situations are possible.

\begin{figure}[htbp]
\centering
\includegraphics[width=0.8\linewidth]{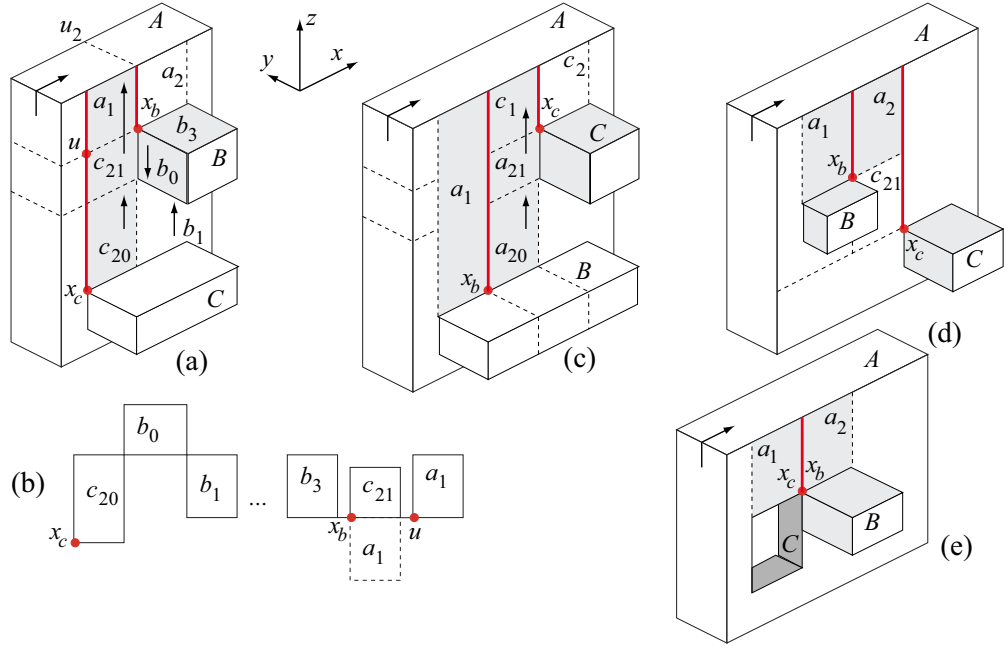}
\caption{(a) Return path for $C$ includes $c_{20}, c_{21}, a_1$;
forward path for $B$ is $a_1$. (b) Unfolding for (a) (c) Return path
for $B$ includes $a_{20}, a_{21}, c_1$; forward path for $C$ is
$c_1$. (d) Return path for $B$ is $a_2$; forward path for $C$
includes $a_2, c_{21}$. (e) Forward (return) paths are identical for
$B$ and $C$.} \label{fig:return}
\end{figure}

\begin{figure}[htbp]
\centering
\includegraphics[width=0.98\linewidth]{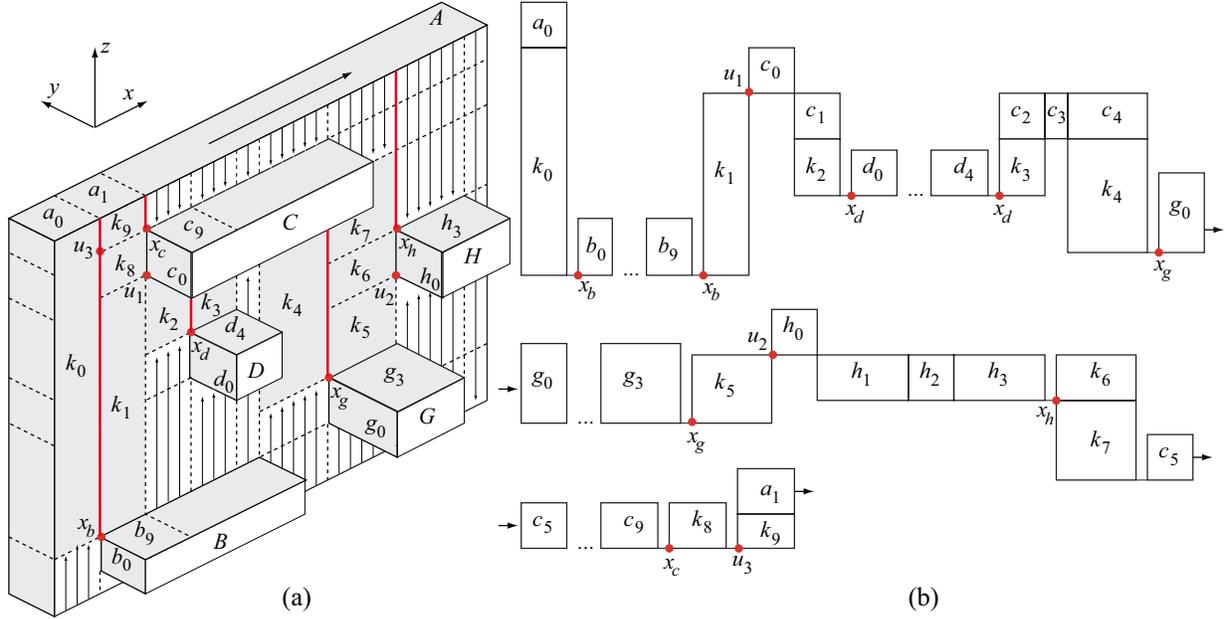}
\caption{(a) An example. (b) The vertex-unfolding.}
\label{fig:ex.1}
\end{figure}

\paragraph{Case 3a:} The forward path $a_1$ for $B$ overlaps the return
path for another descendant $C$ of $A$. This situation is
illustrated in Fig.~\ref{fig:return}a.
In this case, the unfolding $B$ starts as soon as the unfolding
along the return path from $C$ to $A$ meets a face of $B$ incident
to $x_b$ (face $b_0$ in Fig.~\ref{fig:return}a). At this point $B$
gets recursively unfolded as before (see Fig.~\ref{fig:return}b),
then the unfolding continues along the return path for $C$ back to
$A$.
Fig.~\ref{fig:return}b shows face $a_1$ in two positions: we let
$a_1$ hang down only if the next face to unfold is a right face of a
child of $A$ (see the transition from $k_7$ to $c_5$ in
Fig.~\ref{fig:ex.1}); otherwise, use $a_1$ in the upward position, a
freedom permitted to us by rotating about vertex $u$.

\paragraph{Case 3b:}
The return path $a_2$ for $B$ overlaps the forward path for another
descendant $C$ of $A$. This situation is illustrated in
Figs.~\ref{fig:return}c and~\ref{fig:return}d. The case depicted in
Fig.~\ref{fig:return}c is similar to the one in
Fig.~\ref{fig:return}a and is handled in the same manner. For the
case depicted in Fig.~\ref{fig:return}d, notice that $a_2$ is on
both the forward path for $C$ and the return path for $B$. However,
no conflict occurs here: from $a_2$ the unfolding continues downward
along the forward path to $C$ and unfolds $C$ next.

\paragraph{Case 3c:}
The forward path $a_1$ for $B$ overlaps the forward path for another
descendant $C$ of $A$. This situation occurs when either $B$ or
another band $C$ incident to $B$ is a dent, as illustrated in
Figs.~\ref{fig:return}e. Again, no conflict occurs here: the
recursive unfolding of $C$, which returns to $x_c = x_b$, is
followed by the recursive unfolding of $B$, which returns to $x_b$,
then the unfolding continues along the return path for $B$ ($C$)
back to $A$.

\medskip
\noindent
Fig.~\ref{fig:ex.1} shows a more complex example that
emphasizes these subtle unfolding issues.
Note that the return path $k_1, k_8, k_9$ for $B$ overlaps the
forward path $k_9$ for $C$; and the return path
$k_5, k_6$ and $k_7$ for $G$ overlaps the forward path
for $H$, which includes $k_7$.
The unfolding produced by the method described in this section
is depicted in Fig.~\ref{fig:ex.1}(b).

\subsection{Attaching Front and Back Faces to the Net}
\label{sec:faces} Front and back faces of a slab are ``hung'' from
bands following the basic idea of the illumination model discussed
in Sec.~\ref{sec:3x1front.back}. There are three differences,
however, caused by the employment of some front and back gridfaces
for the connecting paths, which can block illumination from the
bands.
\begin{enumerate}
\squeezelist
\item We illuminate both upward and downward from each band:
each $x$-edge illuminates the vertical face it attaches to. This
alone already suffices to handle the example in Fig.~\ref{fig:ex.1}:
all vertical faces are illuminated downward from the top of $A$,
upward from the bottom of $A$, and upward from the top of $B$.

\item Some gridfaces still might not be illuminated by any bands,
because they are obscured both above and below by paths in
connecting faces.  Therefore we incorporate the connecting faces
into the band for the purposes of illumination. For example, in
Fig.~\ref{fig:box-aligned-box}a, $a_2$ illuminates downward and
$a_1$ illuminates upward. The reason this works is that, with one
exception, each vertical connecting strip remains vertical in the
unfolding, and so illuminated strips can be hung safely without
overlap.

\item The one exception is the forward connecting
path $a_1$ in Fig.~\ref{fig:box-aligned-box}b. This paths unfolds
``on its side,'' i.e., what is vertical in 3D becomes horizontal in
2D. Note, however, that the face $x$ below each of these paths (a
face always present), is oriented vertically. We thus consider $x$
to be part of the connecting path for illumination purposes,
permitting the strip below to be hung under $x$.
\end{enumerate}
Because our cases are exhaustive, one can see now that all gridfaces
of (say) the front face of $A$ are either illuminated by $A$, or by
some descendant of $A$ on the front face, augmented by the
connecting paths as just described. (In fact every gridface is
illuminated twice, from above and below.) Hanging the strips then
completes the unfolding.

\subsection{Algorithm Complexity}
Because there are so few unfolding algorithms, that there is
\emph{some} algorithm for a class of objects is more important than
the speed of the algorithm.  Nevertheless, we offer an analysis of
the complexity of our algorithm. Let $n$ be the number of corner
vertices of the polyhedron, and $N=O(n^2)$ the number of gridpoints.
The vertex grid can be easily constructed in $O(N)$ time, leaving a
planar surface map consisting of $O(N)$ gridpoints, gridedges, and
gridfaces. The computation of connecting rays
(Sec.~\ref{sec:spanning.tree}) requires determining the components
of $A \cap P^+$ and $A \cap P-$, for each $A$. This can be easily
read of from the planar map by running through the $n$ vertices of
each of the $O(n)$ bands and determining, for each vertex, whether
it belongs to $P^+$ or $P^-$.
Each of the $O(n)$ band components shoots a vertical ray from one
corner vertex, in a 2D environment (the plane $Y_i$) of $n$
noncrossing orthogonal segments. Determining which band a ray hits
involves a ray shooting query.
%
Although an implementation would employ an efficient data structure,
perhaps BSP trees~\cite{py-obspo-92}, for complexity purposes the
naive $O(n)$ query cost suffices to lead to $O(n^2)$ time to
construct $G_r$. Selecting pivots (Sec.~\ref{sec:pivots}) involves
2-coloring $G_r$ in $O(n)$ time, and computing the unfolding tree
$T_U$ in a breadth-first traversal of $G_r$, which takes $O(n)$
time. Unfolding bands (Sec.~\ref{sec:unfolding.bands}) involves a
depth-first traversal of $T_U$ in $O(n)$ time, and laying out the
$O(N)$ gridfaces in $O(N)$ time. Thus, the algorithm can be
implemented to run in $O(N) = O(n^2)$ time.

\section{Further Work}
Extending these algorithms to arbitrary genus orthogonal polyhedra
remains an interesting open problem. Holes that extend only in the
$x$ and $z$ directions within a slab seem unproblematic, as they
simply disconnect the slab into several components. Holes that
penetrate several slabs (i.e, extend in the $y$ direction) present
new challenges. One idea to handle such holes is to place a virtual
$xz$-face midway through the hole, and treat each half-hole as a
dent (protrusion).

\subsection*{Acknowledgements}
We thank the anonymous referees
on~\cite{dfo-gvuop-06} for their careful reading and insightful
comments.


\bibliographystyle{alpha}
\bibliography{VU} 

\newpage
\section{Appendix: Proof of Lemma~\protect\ref{lem:Gb.connected0} (Connectedness of $G_b$)}
\label{sec:Gb.connected}

Two subsets of $P \subset Y_i$ are \emph{path-connected}, or just
\emph{connected}, if there are points in each that are connected by
a path that lies in $P$. We need some notation to describe the
portions of $r(A)$ that are relevantly connected to each band $A$.
For a protrusion $A$, let $r_c(A)$ be the subset of $r(A)$
(cf.~Sec.~\ref{sec:defs}) that is path-connected to $A$ via paths
that do not cross any bands. For a dent $B$, let $r_c(B)$ be the
boundary of $B$ plus the subset of $r(B)$ that is both
path-connected to $B$ via paths that do not cross any bands, and is
not part of $r_c(A)$, for some protrusion $A$. Consider for example
Figure~\ref{fig:p-d0}b. For protrusion $B'$, $r_c(B')$ consists of
the boundary rim of $B'$ and the portion of the back face of $B'$
that overhangs dent $B$. For dent $B$, $r_c(B)$ consists only the
boundary of $B$, even though the overhanging portion of $B'$ can be
reached from $B$ without crossing any bands, because that is part of
$r_c(B')$. In Figure~\ref{fig:d-d}a however, the portion of the
front face of $A'$ enclosed by $B$ belongs to $r_c(B)$, not to
$r_c(A')$.

The genus-zero assumption implies that, for protrusion $A$ and dent
$B$ on opposite sides of $Y_i$ such that $r_c(A) \cap r_c(B)$ is
nonempty, it must be that $A \cap B$ is nonempty (cf.
Figs.~\ref{fig:p-d}). Define

$$
r_c(A,B) = \left\{
\begin{tabular}{ll}
$A \cap B$,           & if $A \cap B \neq \emptyset$, and at least one of $A$ and $B$ is a dent \\
$r_c(A) \cap r_c(B)$  & otherwise.
\end{tabular}
\right.
$$

This definition is intended to identify gridpoints on either $A$ or
$B$ from which rays are issued by the ray-pair generation algorithm
(Sec.~\ref{sec:ray-pairs}). The reason for treating intersecting
dents and protrusions differently is a subtle one, and is captured
by Fig.~\ref{fig:p-d0}b: $B$ is a dent behind $Y_i$ and $B'$ is a
protrusion in front of $Y_i$; $r_c(B')$ is the piece of the back
face of $B'$ enclosed by $B$; $u$ is a highest gridpoint in $B \cap
B'$, while $w$ is a highest gridpoint in $r_c(B) \cap r_c(B')$; $u$
is a potential ray basepoint, while $w$ is not. The above definition
eliminates points such as $w$ from the set $r_c(A, B)$.

Our connectivity proof for $G_b$ proceeds as follows.  In general,
there are a number of disconnected maximal components $P_1, P_2,
\ldots$ of $P$, with $P = P_1 \cup P_2 \cup \cdots$. The bands
incident to each of these are ray-connected to each other via planes
other than $Y_i$. We first argue that, to prove that $G_b$ is
ray-connected, it suffices to prove that each $P_j$ is
ray-connected. Remove from $O$ all the slabs $S_1, S_2, \ldots$
incident to $Y_0$. Establish that the bands in the resulting object
$O'$ are ray-connected, via induction. Now put back the slabs.  Each
$S_j$ corresponds to a component $P_j$, and we are assuming we can
establish that all bands incident to $P_j$ are ray-connected to one
another.  This along with the fact that $O$ itself is connected
implies that all bands are ray-connected. Henceforth we concentrate
on one such connected component $P_j$, call it $Q \subset Y_i$ for
succinctness. Let $\C$ be the collection of all bands that intersect
$Q$. Then $\displaystyle\cup_{A \in \C} r_c(A) = Q$. The idea of the
connectedness proof is that the bands get connected in upward
chains, and ultimately to each other through ``common ancestor''
higher bands. We choose to prove it by contradiction, arguing that a
highest disconnected component cannot exist.

\begin{lemma}
All bands in $\C$ are ray-connected. Furthermore, if one arbitrary
ray in each ray-pair is discarded, $C$ remains ray-connected.
\label{lem:connected}
\end{lemma}
\begin{proof}
For the purpose of contradiction, assume that not all bands in $\C$
are ray-connected. Let $\C_1, \C_2, \ldots$ be the distinct maximal
subsets of $\C$ that are ray-connected. Let $Q_j = \cup_{A \in \C_j}
r_c(A)$. Then $Q = \cup_j Q_j$. Since $Q$ is connected, the subsets
$Q_j$ are not disjoint, in that for every $Q_j$ there is an $Q_k$
such that $Q_j \cap Q_k$ is nonempty. By the observation above, this
means that
\[
Q_{jk} = \cup_{A \in \C_j, B \in C_k} ~r_c(A, B)
\]
is also nonempty. Let $j$ and $k$ be such that $Q_{jk}$ contains a
{\em highest} $x$-gridedge (gridpoint, if $Q_{jk}$ contains only
isolated points) among all $Q_{jk}$. Let $u$ be the leftmost highest
gridpoint in $Q_{jk}$. Let $A \in \C_j$ and $B \in \C_k$ be such
that $u \in r_c(A, B)$.

We have thus identified two bands $A$ and $B$, ray-disconnected
because in different components of $Q$, which contribute this
highest gridpoint $u$ in the ``highest'' intersection $Q_{jk}$. We
now examine in turn the four  protrusion/dent possibilities for
these two bands.

\medskip
\noindent {\bf Case 1.} $A$ and $B$ are both protrusions on opposite
sides of $Y_i$. Assume w.l.o.g that $A$ is behind $Y_i$, $B$ is in
front of $Y_i$, and $u$ is on $B$ (as depicted in
Fig.~\ref{fig:p-p-opp}).
%
\begin{figure}[htbp]
\centering
\begin{tabular}{c@{\hspace{0.06\linewidth}}c@{\hspace{0.06\linewidth}}c}
\includegraphics[width=0.25\linewidth]{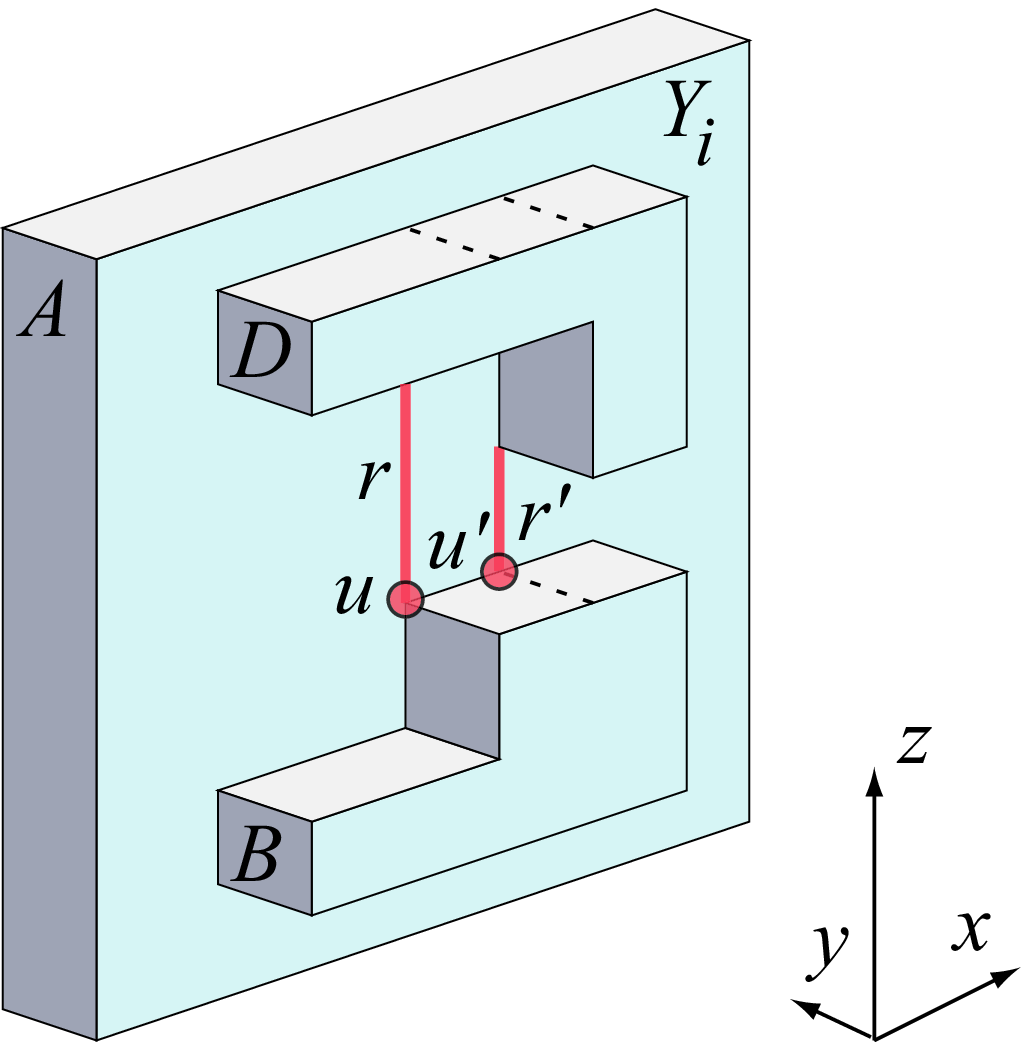} &
\includegraphics[width=0.25\linewidth]{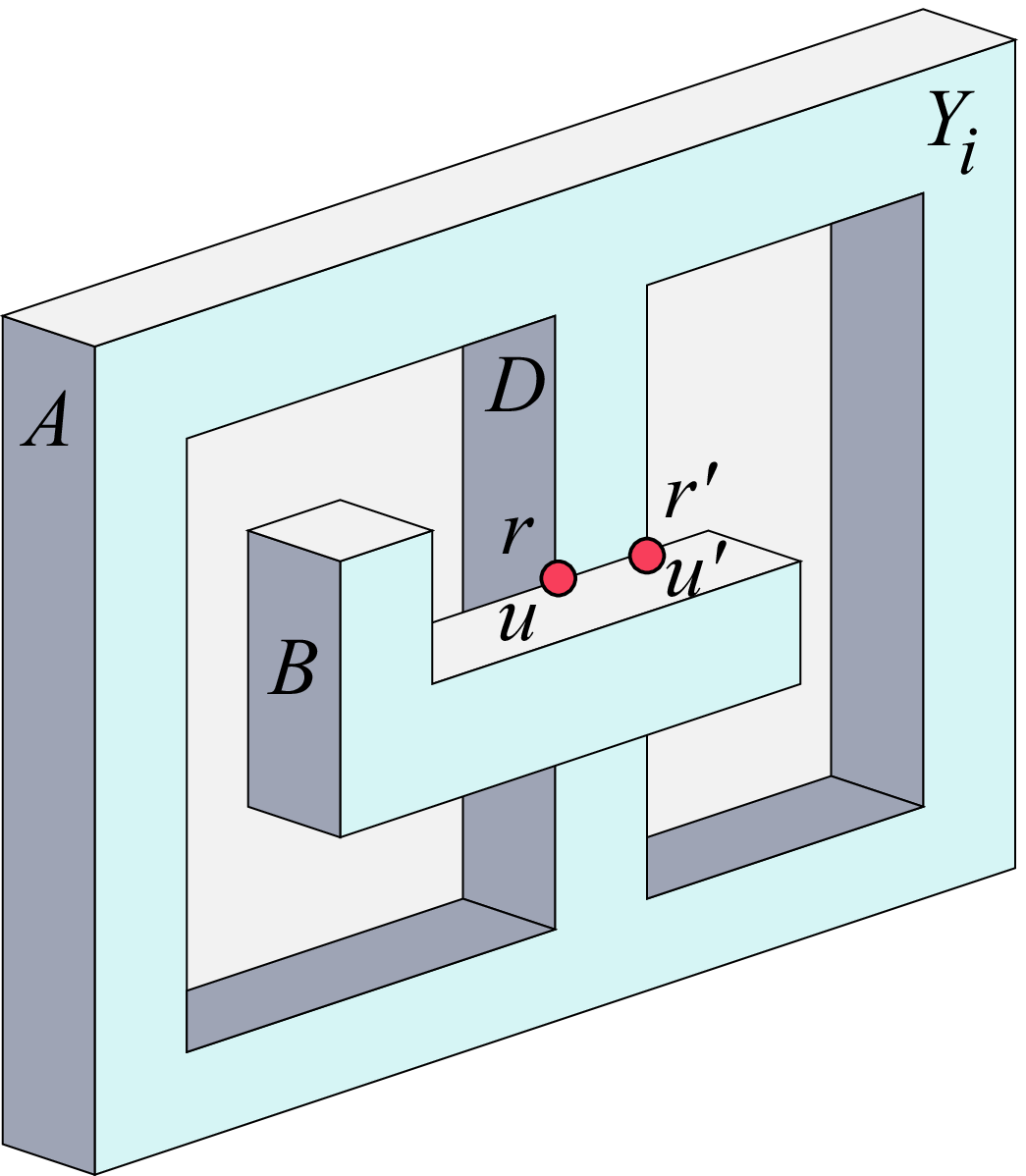} &
\includegraphics[width=0.25\linewidth]{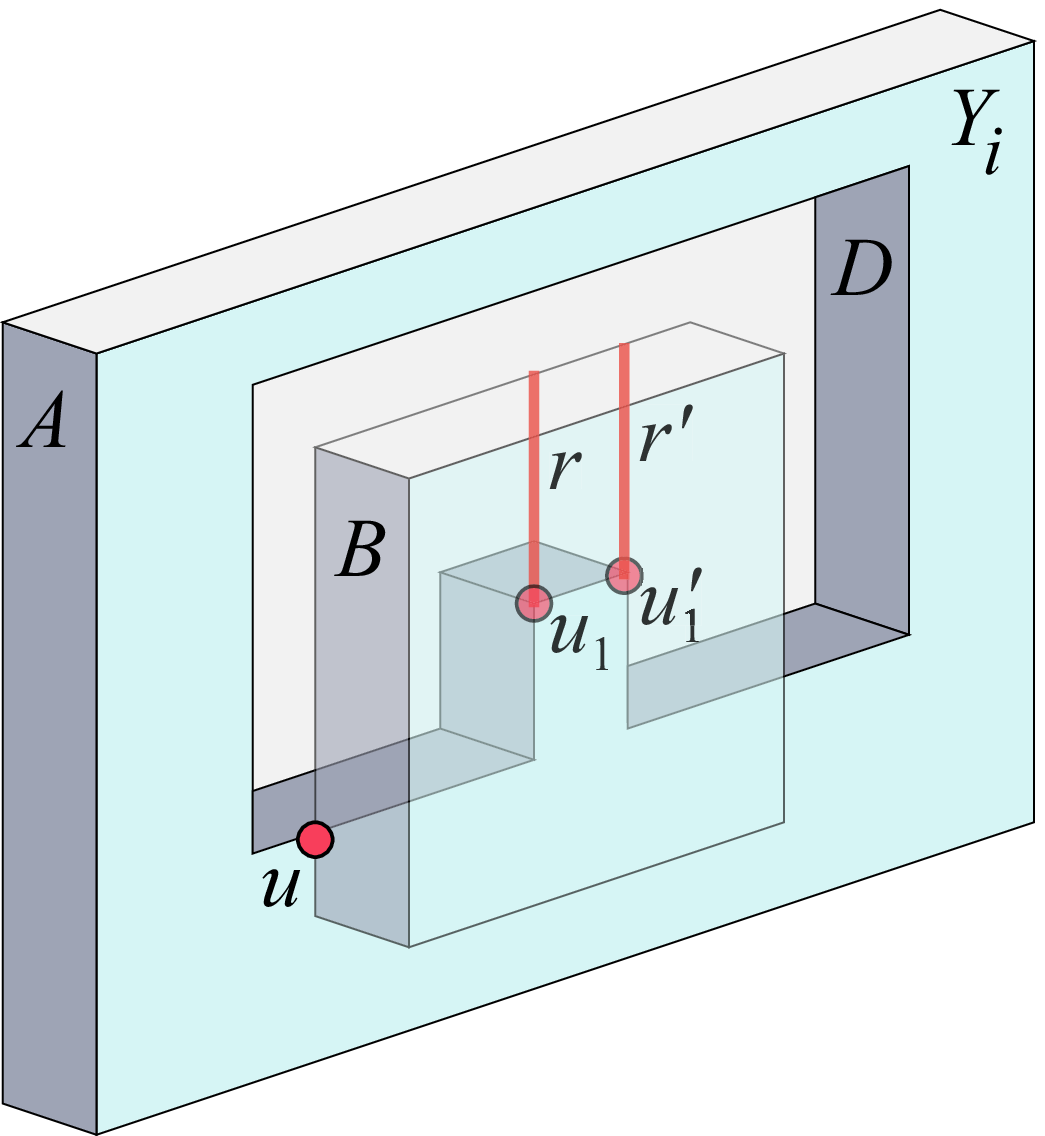} \\
(a) & (b)
\end{tabular}
\caption{Case 1: $A$ and $B$ are both protrusions on opposite sides
of $Y_i$ (a) $D$ is a protrusion (b) $D$ is a dent with a vertical
side incident to $u$ (c) $D$ is a dent with a bottom edge incident
to $u$.} \label{fig:p-p-opp}
\end{figure}
%
We discuss two subcases:
\begin{enumerate}
\item[a.] $u$ is on a top edge of $B$ (Figs.~\ref{fig:p-p-opp}(a,b)).
Then our ray-pair algorithm generates a ray-pair $(r, r')$, with $r$
incident to $u$ and $r'$ incident to the gridpoint $u'$ cw from $u$.
Consider $r$ (the analysis is similar for $r'$). If $r$ hits $A$,
then in fact $A$ and $B$ are ray-connected, contradicting the fact
that $A$ and $B$ belong to different ray-connected components of
$\C$. So let us assume that $r$ hits another band $D \in C_{\ell}$.
Fig.~\ref{fig:p-p-opp}a(b) illustrates the situation when $D$ is a
protrusion (dent). If $D \in \C_j$, then $D$ and $A$ are
ray-connected in $C_j$, and since $B$ and $D$ are ray-connected, it
follows that $B$ and $A$ are ray-connected, a contradiction. So
assume that $D \in \C_{\ell}$, with $\ell \neq j$. But then $r_c(A,
D)$ (and implicitly $Q_{j\ell}$) has a gridpoint higher than $u$,
contradicting our choice of $j$, $k$ and $u$.

\item[b.] $u$ is on a vertical (left, right) edge of $B$
(Fig.~\ref{fig:p-p-opp}c). Then $u$ must be at the intersection
between a dent $D$ and $B$, meaning that $D \cap r_c(B)$ is
nonempty. Furthermore, $r_c(A, D)$ has a gridpoint higher than $u$,
meaning that $D \in G_j$. Let $u_1$ be the leftmost among the
highest gridpoints of $D \cap r_c(B)$. Then our ray-pair algorithm
generates a ray-pair $(r, r')$ from $u_1$ and its right neighbor
$u'_1$. Consider $r$ (the analysis is similar for $r'$). If $r$ hits
$B$, then $B$ is ray-connected to $D$, which is ray-connected to
$A$, a contradiction. If $r$ hits a band $E$ other than $D$, then it
must be that $D \in C_k$, since $r_c(B,E)$ has a gridpoint higher
than $u_1$, which is no lower than $u$. This means that $B$ is
ray-connected to $E$, which is ray-connected to $D$, which is
ray-connected to $A$, a contradiction.
\end{enumerate}

\medskip
\noindent {\bf Case 2.} $A$ is a protrusion and $B$ is a dent, both
on a same side of $Y_i$. The case when $A$ and $B$ are both in
front of $Y_i$ (illustrated in Fig.~\ref{fig:p-d0}a) is identical to
Case 1 above, once one conceptually pops out $B$ into a protrusion.
We now discuss the case when $A$ and $B$ are both behind $Y_i$.

Assume first that $r_c(A, B)$ contains no top edges of $B$, as
depicted in Figure~\ref{fig:p-d0}b. Let $B'$ be a protrusion in
front of $Y_i$ covering the top of $B$. Then $r_c(A,B')$ and
$r_c(B',B)$ each contains a gridpoint higher than $u$. The following
two contradictory observations settle this case:
\begin{enumerate}
\item[a.] It must be that $B' \not\in \C_k$; otherwise $Q_{jk}$ would contain
a gridpoint in $r_c(A,B')$ higher than $u$.

\item[b.] If $B' \in C_{\ell}$, then it must be that $\ell = k$; otherwise
$Q_{\ell k}$ would contain a gridpoint in $r_c(B',B)$ higher than
$u$.
\end{enumerate}
If $r_c(A,B)$ contains at least one top gridedge of $B$, then
arguments similar to the ones used for the case illustrated in
Fig.~\ref{fig:p-p-opp}a (conceptually popping $B$ to become a
protrusion) settle this case as well.
%
\vspace{-1em}
\begin{figure}[htbp]
\centering
\begin{tabular}{c@{\hspace{0.15\linewidth}}c}
\includegraphics[width=0.32\linewidth]{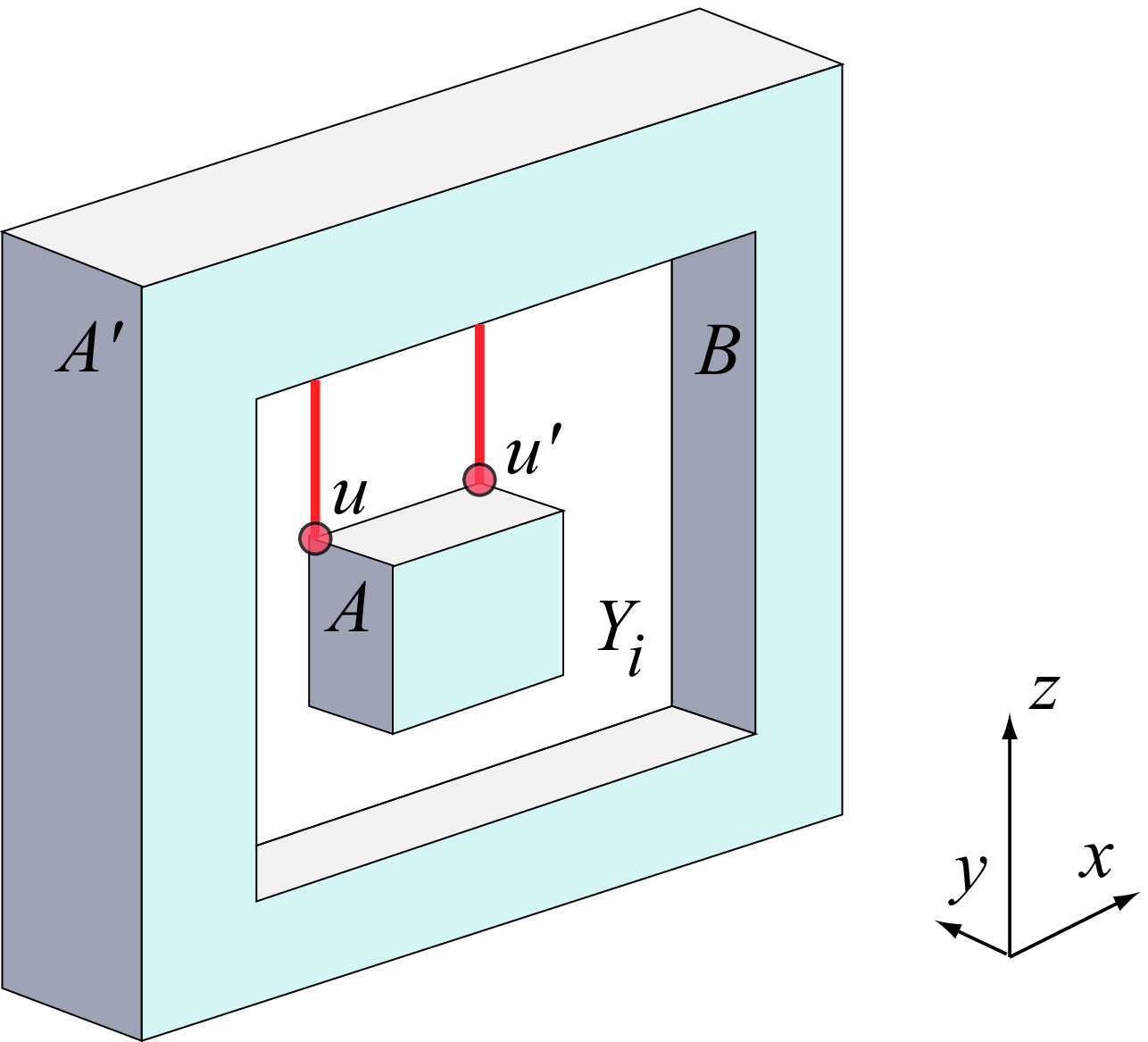} &
\includegraphics[width=0.24\linewidth]{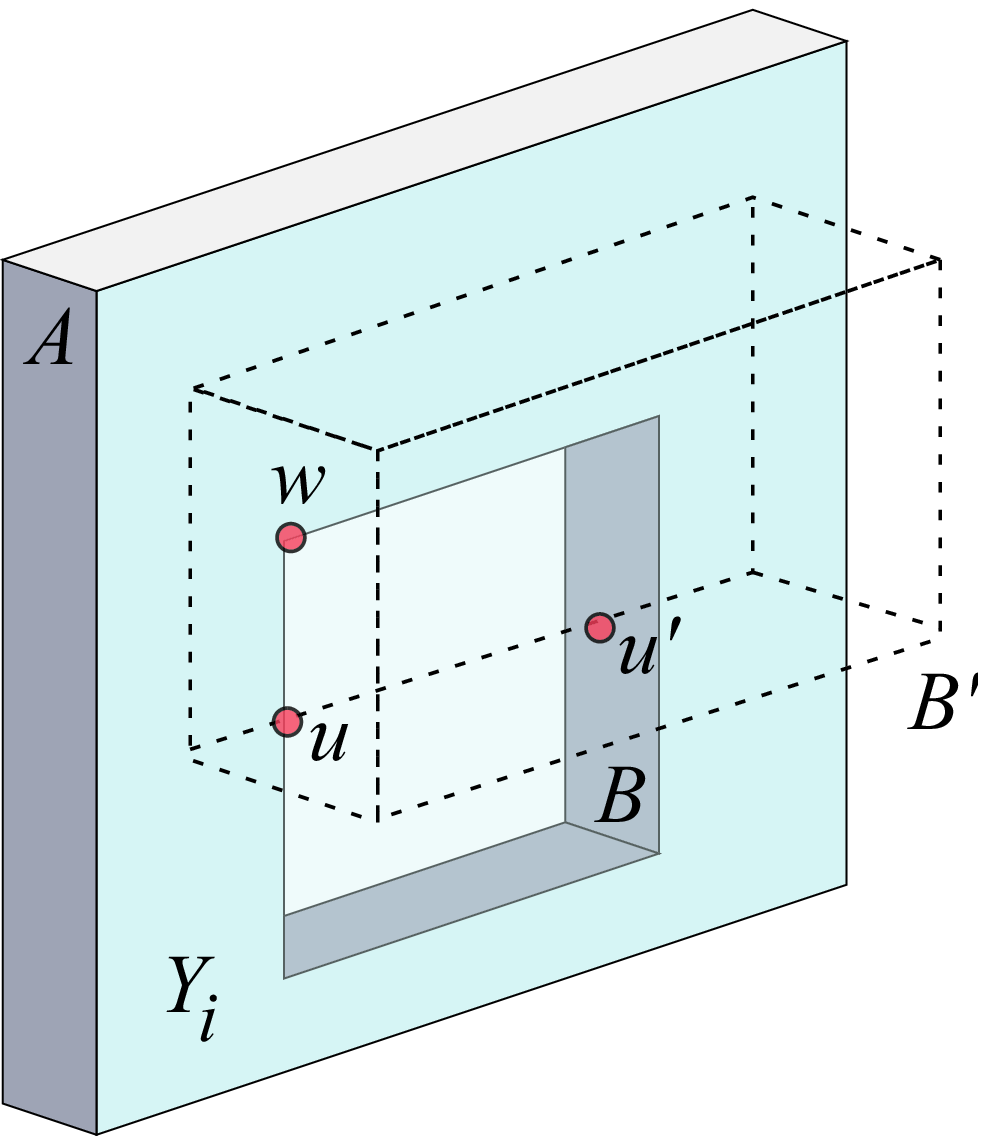} \\
(a) & (b)
\end{tabular}
\caption{Case 2: $A$ is a protrusion and $B$ is a dent (a) behind
$Y_i$ (b) in front of $Y_i$.} \label{fig:p-d0}
\end{figure}

\medskip
\noindent {\bf Case 3.} $A$ is a protrusion and $B$ is a dent on
opposite sides of $Y_i$ (see Fig.~\ref{fig:p-d}). Let $B'$ be the
protrusion in front of $Y_i$ enclosing $B$.
We discuss two subcases:
\begin{enumerate}
\item[a.] $r_c(A)$ contains a top edge of $B$
(see Fig.~\ref{fig:p-d}a). This means that $r_c(A) \cap r(B)$ is
nonempty, and the ray-pair algorithm shoots a ray-pair $(r, r')$
upward from the endpoints of a highest gridedge $\{u_1, u'_1\}$ of
$A \cap r(B)$. Consider ray $r$ (the analysis is similar for $r')$.
If $r$ hits $B$, then $A$ and $B$ are in fact ray-connected, a
contradiction. If $r$ hits a band $D$ other than $B$, then arguments
similar to the ones for the case illustrated in
Fig.~\ref{fig:p-p-opp}a (Case 1) lead to a contradiction.
%
\vspace{-1em}
\begin{figure}[htbp]
\centering
\begin{tabular}{c@{\hspace{0.15\linewidth}}c}
\includegraphics[width=0.25\linewidth]{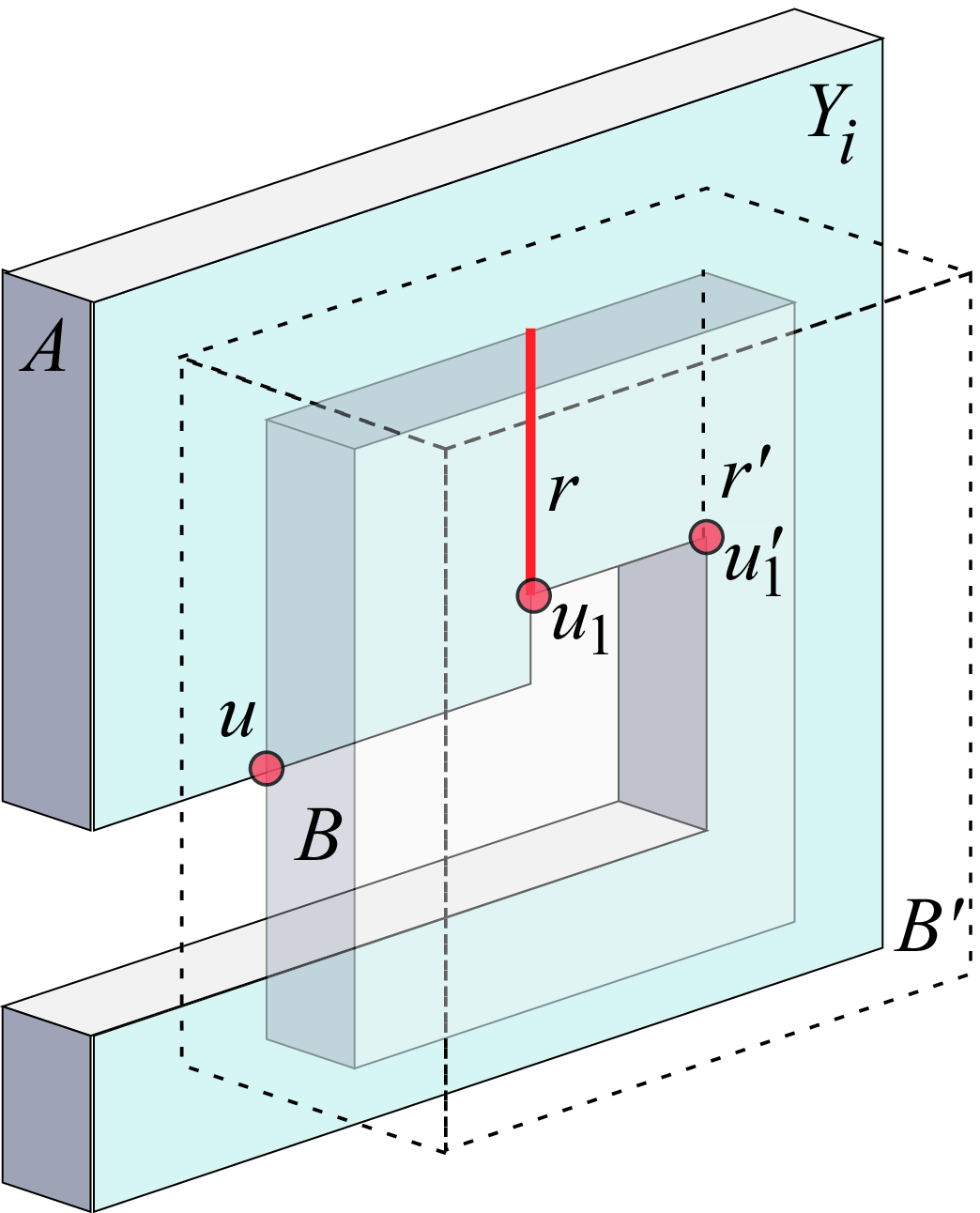} &
\includegraphics[width=0.25\linewidth]{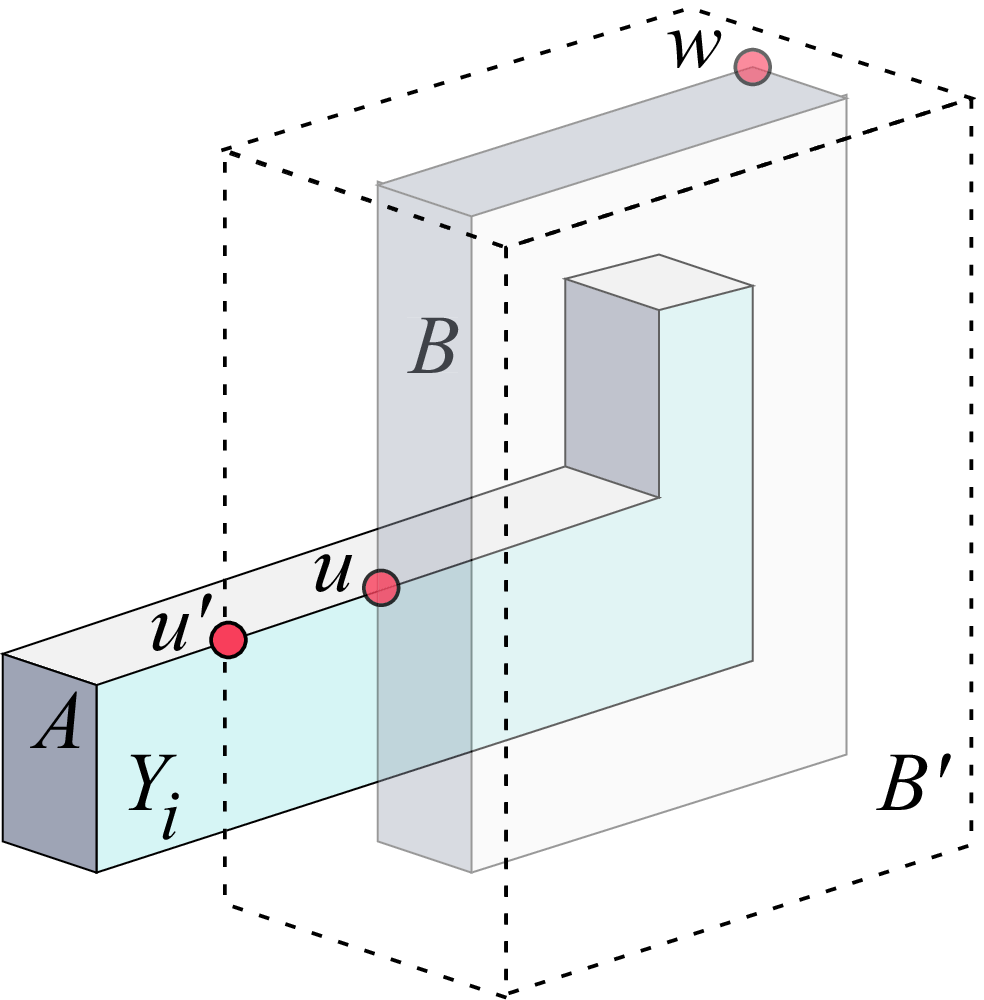} \\
(a) & (b)
\end{tabular}
\caption{Case 3: $A$ is a protrusion behind $Y_i$; $B$ is a dent in
$B'$, both in front of $Y_i$.} \label{fig:p-d}
\end{figure}
%

\item[b.] $r_c(A)$ contains a bottom edge of $B$.
This case is symmetrical to the one above in that a ray upward from
a gridpoint of $B \cap r(A)$ hits $A$, thus ray-connecting $A$ and
$B$.
\item[c.] $r_c(A)$ contains neither a top nor a bottom
edge of $B$ (see Fig.~\ref{fig:p-d}b). Arguments similar to the ones
used in Case 1 (protrusions on opposite sides of $Y_i)$ show that
$A$ and $B'$ are ray-connected. That $B$ and $B'$ are ray-connected
follows immediately from the fact that $r_c(B, B')$ has a gridpoint
higher than $u$ ($w$ in Fig.~\ref{fig:p-d}b). These together imply
that $A$ and $B$ are ray-connected, a contradiction.
\end{enumerate}
%
\begin{figure}[htbp]
\centering
\begin{tabular}{c@{\hspace{0.15\linewidth}}c}
\includegraphics[width=0.32\linewidth]{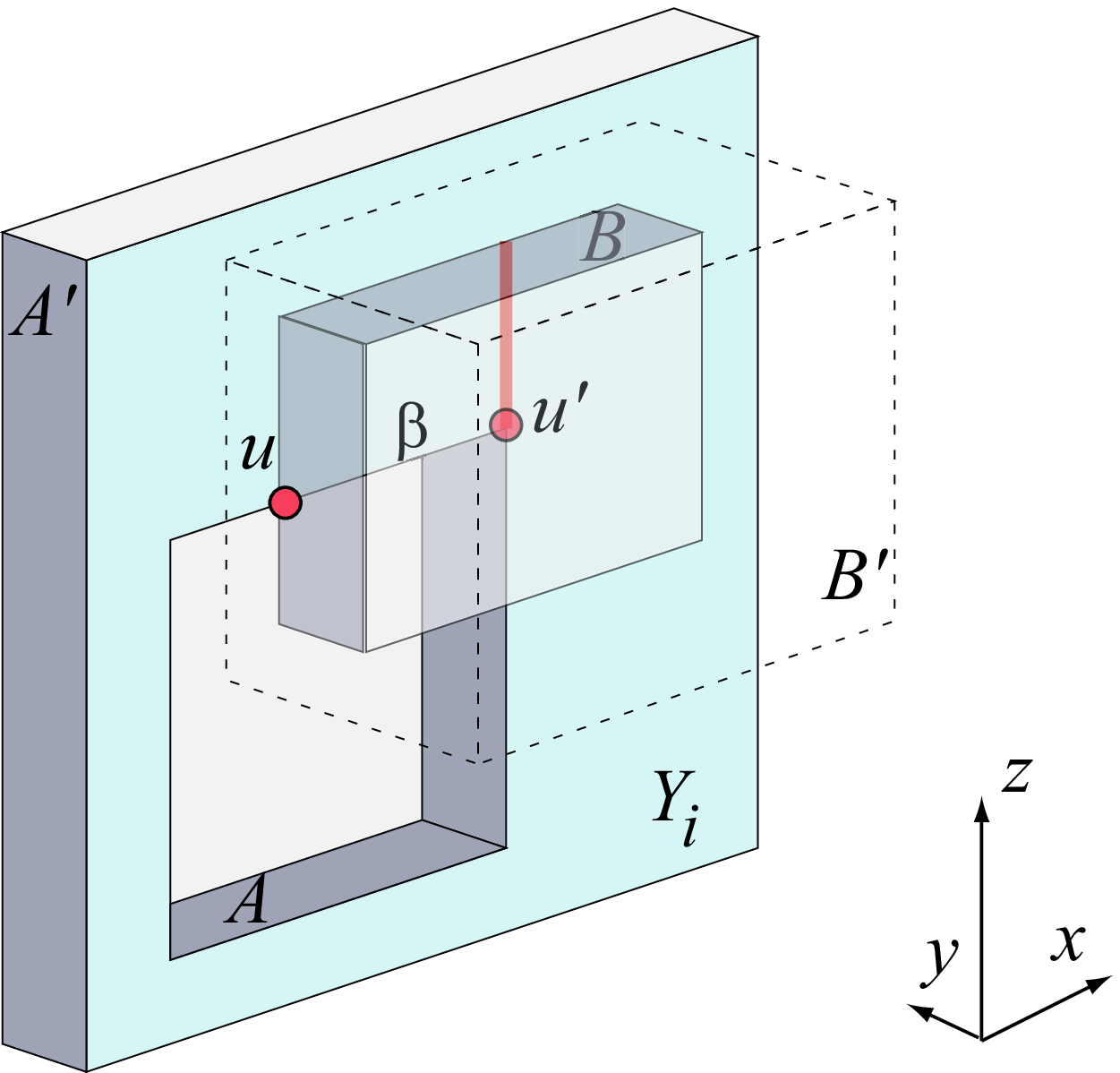} &
\includegraphics[width=0.25\linewidth]{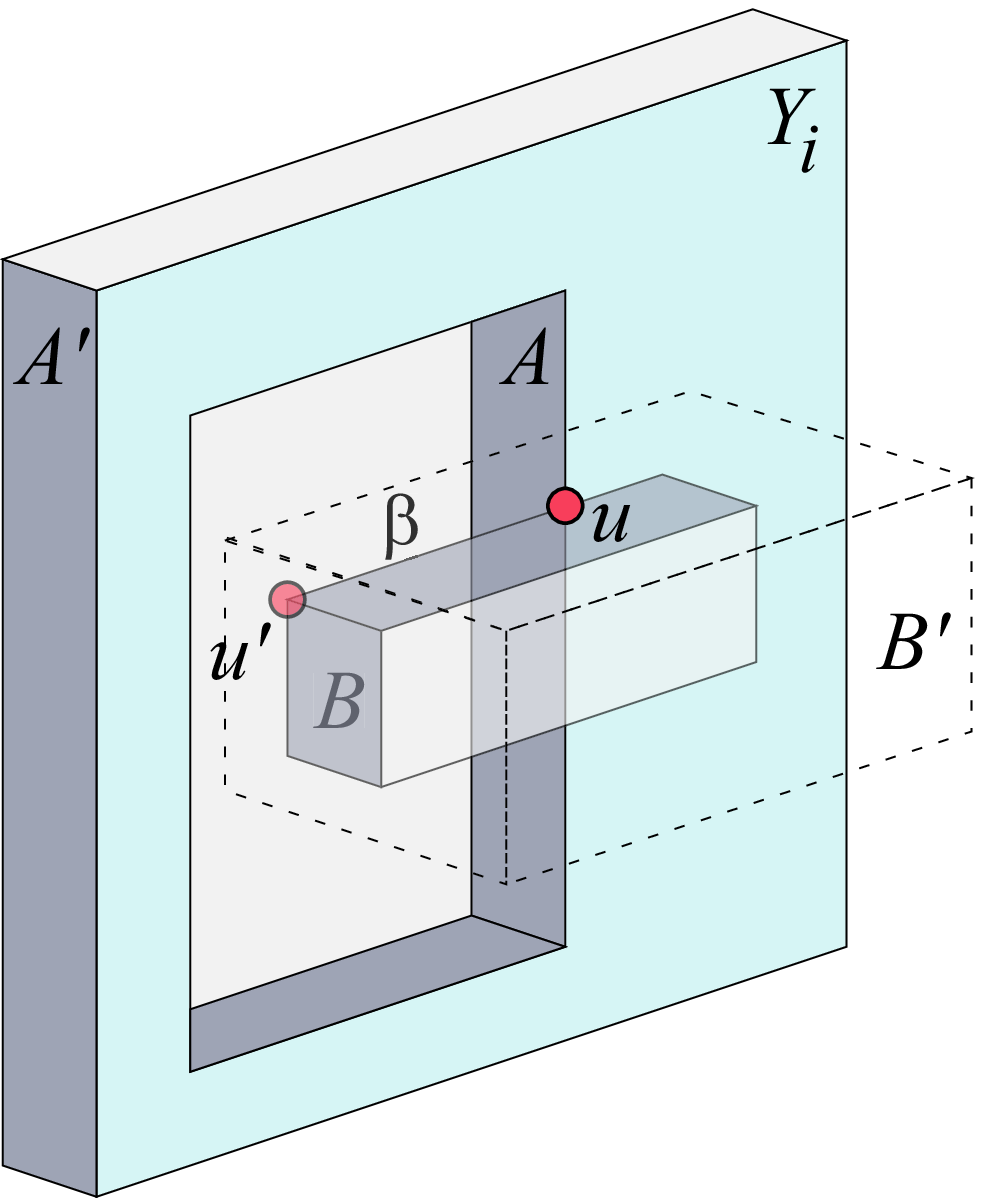} \\
(a) & (b)
\end{tabular}
\caption{Case 4: $A$ is a dent behind $Y_i$, enclosed within
protrusion $A'$. $B$ is a dent in front of $Y_i$, enclosed within
protrusion $B'$.} \label{fig:d-d}
\end{figure}
%

\medskip
\noindent {\bf Case 4.} $A$ and $B$ are both dents: $A$ is a dent
behind $Y_i$ enclosed within protrusion $A'$, and $B$ is a dent in
front of $Y_i$ enclosed within protrusion $B'$ (see
Fig.~\ref{fig:d-d}). The genus-zero assumption implies that $r(A)
\cap r(B)$ is a polygonal region of positive area. Since $u \in
r_c(A) \cap r_c(B)$, we have that $u \in r(A) \cap r(B)$. Let
$\beta$ be the boundary segment of $r(A) \cap r(B)$ incident to $u$.
We discuss two subcases:
\begin{enumerate}
\item[a.] $\beta \subset P^-$, meaning that $\beta \subset A$
(see Fig.~\ref{fig:d-d}a).

An analysis similar to the one for the case illustrated in
Fig.~\ref{fig:p-d}a (Case 3) shows that $A$ and $B$ are
ray-connected, a contradiction.

\item[b.] $\beta \subset P^+$, meaning that $\beta \subset B$
(see Fig.~\ref{fig:d-d}b). We show that $A$ and $A'$ are
ray-connected, $B$ and $B'$ are ray-connected, and $A'$ and $B'$ are
ray-connected. This implies that $A$ and $B$ are ray-connected, a
contradiction. First note that the ray-pair algorithm shoots a
ray-pair $(r, r')$ upward from a highest gridedge on $\beta$. An
analysis similar to the one for the case illustrated in
Fig.~\ref{fig:p-p-opp}a (conceptually popping $B$ to become a
protrusion) shows that $r$ and $r'$ must hit $B$, thus
ray-connecting $B$ and $B'$. That $A$ and $A'$ are ray-connected
follows immediately from the fact that $r_c(A, A')$ has a gridpoint
higher than $u$, and similarly for $A'$ and $B'$.
\end{enumerate}
Having exhausted all possible cases, the connectivity claim of the
lemma is established. Because the proof for each of these cases goes
through by considering either the first or second ray of a ray-pair,
retaining either ray suffices to preserve connectivity. Thus the
second claim of the lemma is established as well.
\end{proof}

\end{document}